\documentclass[aps,pre,twocolumn,groupedaddress,showpacs,superscriptaddress]{revtex4-1}
\usepackage{graphicx}
\usepackage{wasysym}
\usepackage{amssymb}
\usepackage{color}
\usepackage{tikz}
\usepackage{pgfplots}
\usetikzlibrary{shapes,arrows,positioning,shadows,decorations.pathreplacing,patterns,plotmarks}

\begin{document}
\title{Global space-time update}

\author{V.G.~Rousseau}
\affiliation{Department of Physics and Astronomy, Louisiana State University, Baton Rouge, Louisiana 70803, USA}
\author{D.~Galanakis}
\affiliation{School of Physical and Mathematical Sciences, Nanyang Technological University, Singapore 637371}
\relpenalty=10000      
\binoppenalty=10000
\begin{abstract}
The Stochastic Green Function (SGF) algorithm is able to simulate any Hamiltonian that does not suffer from the so-called
``sign problem". We propose a new global space-time update scheme for the SGF algorithm which, in addition to being simpler than
the previous formulation, reduces auto-correlation times.
Using as a concrete example the extended Bose-Hubbard model and the complex Hamiltonian with six-site ring-exchange
interactions which was recently studied in ArXiv:1206.2566v1, we present a comprehensive review of the SGF algorithm
and the new updating scheme. Measurements of non-trivial physical quantities are presented in detail. While the SGF algorithm
works in the canonical ensemble by nature, we give a simple extension that allows to perform simulations in the grand-canonical
ensemble too. We also discuss an optimized implementation which allows for access to large system sizes.
\end{abstract}

\maketitle

\section{Introduction}
Monte Carlo methods were first introduced more than 60 years ago~\cite{Metropolis} to solve various classical problems.
The development of quantum Monte Carlo (QMC)~\cite{Handscomb,Kalos,Blankenbecler} allowed the extension of those methods to quantum systems.
For fermions, the power of QMC methods remains limited by the so-called \textit{sign problem}, which prevents the treatment of
systems with large sizes. While the Determinant Quantum Monte Carlo algorithm~\cite{Blankenbecler} allows to treat fermions exactly in
some particular cases, it is often necessary to make use of approximate methods such as the fixed-node approximation~\cite{Anderson}, and
the dynamical cluster approximation~\cite{Hettler1,Hettler2}.
On the other hand there are many interesting bosonic systems which do not suffer from the sign problem, and which
can be described efficiently using a worldline representation~\cite{Batrouni}.
Consequently, over the last two decades, there have been tremendous advances in boson QMC methods,
such as the development of the Stochastic Series Expansion (SSE) algorithm~\cite{Sandvik},
the loop algorithm~\cite{Evertz} which makes use of global updates in the grand-canonical ensemble, or the Reptation Quantum Monte Carlo algorithm~\cite{Carleo}.
In recent years the canonical worm (CW) algorithm~\cite{VanHoucke}, which
works in the canonical ensemble with global updates, was proposed. The method was generalized later to a wider class of
Hamiltonians, leading to the Stochastic Green Function (SGF) algorithm~\cite{SGF}, and improved with a new type of \textit{directed} updates~\cite{DirectedSGF}.

Recently, the generality of the SGF algorithm made it possible to perform exact studies of complex systems, such as multispecies systems with
interspecies conversions~\cite{Feshbach,Spin1Bosons,SpinHalfBosons1,SpinHalfBosons2,SpinHalfBosons3}, systems with fully connected
graphs~\cite{Carleo2}, and systems described by Hamiltonians with six-site coupling terms~\cite{RousseauArXiv}.

In this paper we review the theory of the SGF algorithm and propose a new \textit{global space-time update} with reduced
auto-correlation times.
The paper is organized as follows:
In section II we describe the properties and the framework of the SGF algorithm.
In section III we present the new global space-time update scheme and derive the expression of all associated probabilities that
satisfy detailed balance. We detail the differences between this new update and the updates that are used in the SSE and CW algorithms.
Section IV gives full details on measurements.
In particular we describe how non-trivial quantities can be measured, such as the specific heat, the imaginary dynamical structure factor,
the entropy, or $n$-point Green functions.
We give in section V a simple extension that allows the algorithm to
simulate exactly the grand-canonical ensemble. We propose in Section VI an efficient implementation of the algorithm. We illustrate
in section VII the exactness of the algorithm by making comparisons between the SGF algorithm and exact diagonalizations, and show
that the new global space-time update leads to smaller auto-correlation times. Finally we conclude in section VIII.

\section{Generalities}
\subsection{Properties}
The SGF algorithm is characterized by the following properties:
\begin{enumerate}
  \item It can be applied to any sign-problem-free Hamiltonian.
  \item It is completely independent of the structure of the Hamiltonian, no particular decomposition is required.
  \item As a corollary, it is possible to write a single computer code that can simulate any sign-problem-free Hamiltonian.
  \item The acceptance rate of every update is 100\%. The benefit of that is efficiency, as no cpu time is wasted
  with useless rejected updates and implementation simplicity because the changes made in the configuration
  during an update do not need to be stored.
  \item It can simulate both the canonical and the grand-canonical ensembles.
  \item It makes use of a global space-time update (see section~\ref{GlobalSpaceTimeUpdate}), which reduces the auto-correlation time (see section~\ref{Autocorrelation}).
  \item It allows for the measurement of high-order correlation functions, such as $n$-point Green functions.
  \item It works in continuous imaginary time and is exact (no errors beyond statistical errors).
  \item It ensures the ergodicity. In particular, the winding is sampled.
\end{enumerate}
The above properties $1-7$ illustrate the differences between the SGF algorithm and other existing algorithms. In particular,
while other algorithms can treat only a specific class of Hamiltonians (usually Hamiltonians that can be written as
a sum of two-site coupling terms), the SGF algorithm can be directly applied to any sign-problem-free Hamiltonian ``as is".

In order to give a concrete illustration of the above abilities of the SGF algorithm, we consider in the following two Hamiltonians.
The first Hamiltonian is the one-dimensional extended Bose-Hubbard model, which is familiar:
\begin{eqnarray}
  \nonumber \hat\mathcal H &=& -t\sum_{\big\langle i,j\big\rangle}\big(a_i^\dagger a_j^{\phantom\dagger}+H.c.\big)+\frac{U}{2}\sum_i\hat n_i(\hat n_i-1)\\
  \label{BoseHubbard}      &&  +V\sum_{\big\langle i,j\big\rangle}\hat n_i\hat n_j
\end{eqnarray}
The operators $a_i^\dagger$ and $a_i^{\phantom\dagger}$ are the creation and annihilation operators of a boson on site $i$, which satisfy the usual
boson commutation rules, and $\hat n_i=a_i^\dagger a_i^{\phantom\dagger}$ is the number operator on site $i$. The sum $\sum_{\langle i,j\rangle}$ is over all pairs of first-neighboring sites $i$ and $j$.
The second Hamiltonian describes soft-core bosons on a two-dimensional kagome lattice interacting via an onsite repulsion potential and
a sextic ring-exchange term. The Hamiltonian takes the form:
\begin{eqnarray}
  \nonumber \hat\mathcal H=  &-& t\sum_{\langle i,j\rangle}\big(a_i^\dagger a_j^{\phantom\dagger}+H.c.\big)\\
  \nonumber                  &-& K\sum_{\hexagon}\big(a_{\hexagon 1}^\dagger a_{\hexagon 3}^\dagger a_{\hexagon 5}^\dagger a_{\hexagon 6}^{\phantom\dagger} a_{\hexagon 4}^{\phantom\dagger} a_{\hexagon 2}^{\phantom\dagger}+H.c.\big)\\
  \label{HamiltonianExample} &+& U\sum_{i}\hat n_i\big(\hat n_i-1\big)
\end{eqnarray}
The sum $\sum_{\hexagon}$ is over all hexagons of the lattice. The indices $\hexagon 1-\hexagon 6$ in the sextic term denote the sites of a given hexagon~$\hexagon$.
This Hamiltonian has been studied recently in reference~\cite{RousseauArXiv} in the hard-core case. Because of the sextic term which couples
six sites at a time, simulating this Hamiltonian with usual methods is cumbersome. We demonstrate below that
it is straightforward to simulate it with the SGF algorithm.

\subsection{The (extended) partition function and the Green operator}
Like many other QMC algorithms, the SGF algorithm operates directly on physical states. An occupation number basis
is chosen, $\mathcal B=\big\lbrace\big|\psi\big\rangle\big\rbrace$, where $\big|\psi\big\rangle$ is a configuration in the occupation number representation.
We consider a Hamiltonian written in the form
\begin{equation}
  \hat\mathcal H=\hat\mathcal V-\hat\mathcal T,
\end{equation}
where $\hat\mathcal V$ is diagonal in the basis $\mathcal B$, and $\hat\mathcal T$ is off-diagonal and assumed to have positive matrix elements.
The SGF algorithm does not require any further assumptions on the Hamiltonian.
Defining the inverse temperature $\beta$, the purpose of the algorithm is to sample the partition function
\begin{equation}
  \label{PartitionFunction} \mathcal Z(\beta)=\textrm{Tr }e^{-\beta\hat\mathcal H}.
\end{equation}
To this end, we define the \textit{Green operator} $\hat\mathcal G$ by its matrix elements for all pairs of states $\big|L\big\rangle$ and
$\big|R\big\rangle$,
\begin{equation}
  \label{GreenOperator} \big\langle L\big|\hat\mathcal G\big|R\big\rangle=g(n),
\end{equation}
where $n$ is the number of creations and annihilations needed to transform the state $\big|R\big\rangle$ into the state $\big|L\big\rangle$,
and $g$ is an arbitrary function (see section III), with the constraint $g(0)=1$, that is all diagonal matrix elements of $\hat\mathcal G$
are equal to $1$. In the following, $n$ will be referred to as the
\textit{offset} of the Green operator. By breaking up the exponential in (\ref{PartitionFunction}) at imaginary time $\tau$ and introducing the
Green operator between the two parts, we can define an \textit{extended} partition function:
\begin{equation}
  \label{ExtendedPartitionFunction1} \mathcal Z(\beta,\tau)=\textrm{Tr }e^{-(\beta-\tau)\hat\mathcal H}\hat\mathcal G e^{-\tau\hat\mathcal H}
\end{equation}

Defining $\hat\mathcal T(\tau)=e^{\tau\hat\mathcal V}\hat\mathcal Te^{-\tau\hat\mathcal V}$ and $\hat\mathcal G(\tau)=e^{\tau\hat\mathcal V}\hat\mathcal Ge^{-\tau\hat\mathcal V}$,
and using the equality
\begin{equation}
  \label{InteractionPicture} e^{-\beta(\hat\mathcal V-\hat\mathcal T)}=e^{-\beta\hat\mathcal V}\stackrel{\leftarrow}{\textrm{\bf T}_{\tau^\prime}}e^{\int_0^\beta\hat\mathcal T(\tau^\prime)d\tau^\prime},
\end{equation}
where $\stackrel{\leftarrow}{\textrm{\bf T}_{\tau^\prime}}$ denotes the time-ordering operator over the variable $\tau^\prime$ with
time increasing from the right to the left, the extended partition function takes the form:
\begin{equation}
  \label{ExtendedPartitionFunction2} \mathcal Z(\beta,\tau)=\textrm{Tr }e^{-\beta\hat\mathcal V}\stackrel{\leftarrow}{\textrm{\bf T}_{\tau^\prime}}\Big[e^{\int_\tau^\beta\hat\mathcal T(\tau^\prime)d\tau^\prime}\hat\mathcal G(\tau)e^{\int_0^\tau\hat\mathcal T(\tau^\prime)d\tau^\prime}\Big]
\end{equation}
By expanding the exponentials in (\ref{ExtendedPartitionFunction2}) and ordering the operators in imaginary time, we get:
\begin{eqnarray}
  \nonumber                          \mathcal Z(\beta,\tau) &=&      \textrm{Tr }e^{-\beta\hat\mathcal V}\sum_{n\geq 0}\int_{0<\tau_1<\cdots<\tau_n<\beta}\hspace{-2.0cm}\hat\mathcal T(\tau_n)\cdots\hat\mathcal T(\tau_{L_2})\hat\mathcal T(\tau_{L_1})\hat\mathcal T(\tau_L)\hat\mathcal G(\tau)\\
  \label{ExtendedPartitionFunction3}                        &\times& \hat\mathcal T(\tau_R)\hat\mathcal T(\tau_{R_1})\hat\mathcal T(\tau_{R_2})\cdots\hat\mathcal T(\tau_1)d\tau_1\cdots d\tau_n
\end{eqnarray}
Note that we have used the labels $L$, $L_1$, $L_2$, ... (resp. $R$, $R_1$, $R_2$, ...) for the first, second, third, ... time indices
that appear on the left (resp. right) of $\hat\mathcal G(\tau)$. By introducing $n$ complete sets of states $\sum_{k}\big|k\big\rangle\big\langle k\big|$
with $k=1\cdots n$ between each $\hat\mathcal T$ and $\hat\mathcal G$ operators, and an extra set with $k=0$ for the trace, the extended partition function
takes the final form
\begin{eqnarray}
  \nonumber\mathcal Z(\beta,\tau) &=& \sum_{n\geq 0}\sum_{\big\lbrace\big|k\big\rangle\big\rbrace}\!\!e^{-\beta V_0}\!\!\int_{0<\tau_1<\cdots<\tau_n<\beta}\hspace{-1.9cm}\big\langle 0\big|\hat\mathcal T(\tau_n)\big|n\big\rangle \cdots \\
  \nonumber &\times& \big\langle L_1\big|\hat\mathcal T(\tau_L)\big|L\big\rangle \big\langle L\big|\hat\mathcal G(\tau)\big|R\big\rangle\big\langle R\big|\hat\mathcal T(\tau_R)\big| R_1\big\rangle\\
  \label{ExtendedPartitionFunction4} &\times& \cdots \big\langle 1\big|\hat\mathcal T(\tau_1)\big|0\big\rangle d\tau_1\cdots d\tau_n,
\end{eqnarray}
where we have used the notation $V_k$ (here with $k=0$) to denote the matrix element $\big\langle k\big|\hat\mathcal V\big|k\big\rangle$.
As a result, a configuration of the extended partition function~(\ref{ExtendedPartitionFunction4}) is determined by a set of $n$ time
indices and $n+1$ states~$\big|k\big\rangle$. Since the states~$\big|k\big\rangle$ are connected to each other by single $\hat\mathcal T$ operators,
it is actually simpler to generate the set of $n+1$ states starting from the two states $\big|L\big\rangle$ and $\big|R\big\rangle$ and a set of $n$
particular terms $\hat\mathcal T_k$ of the operator $\hat\mathcal T$. In the following, a configuration for which $\big|L\big\rangle=\big|R\big\rangle$
will be referred to as a \textit{diagonal configuration}, and a particular term $\hat\mathcal T_k$ will be called
\textit{term with space index $k$}.

\begin{figure}[h]
  \centerline{\includegraphics[width=0.45\textwidth]{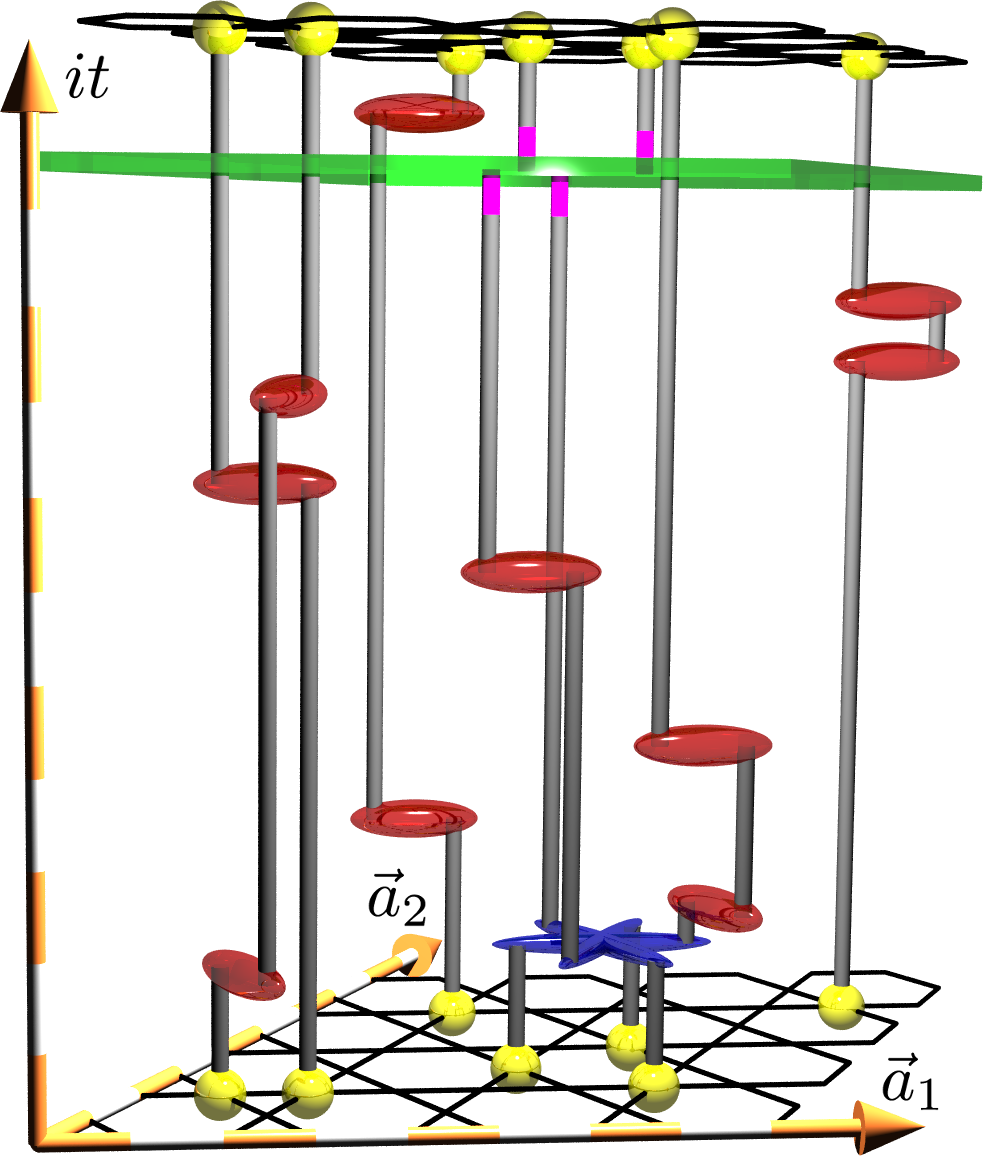}}
  \caption{(Color online) A possible configuration of the extended partition function $\mathcal Z(\beta,\tau)$ for the Hamiltonian~(\ref{HamiltonianExample}). The kagome lattice is represented
  in black with its primary directions $\vec a_1$ and $\vec a_2$ at imaginary times $0$ and $\beta$. The particles (yellow) represent the
  current state of the trace (same state at times $0$ and $\beta$). Some kinetic (red) and ring-exchange
  (blue) terms are distributed in both space and time. The Green operator (green) reconnects the broken worldlines (gray). In the
  present case, the offset (purple) is $n=4$.}
  \label{Configuration}
\end{figure}

In our example~(\ref{HamiltonianExample}), the $\hat\mathcal V$ and $\hat\mathcal T$ operators are identified as
\begin{eqnarray}
  \hat\mathcal V&=& U\sum_{i}\hat n_i\big(\hat n_i-1\big),\\
  \nonumber \hat\mathcal T &=& t\sum_{\langle i,j\rangle}\big(a_i^\dagger a_j^{\phantom\dagger}+H.c.\big)\\
                           &+& K\sum_{\hexagon}\big(a_{\hexagon 1}^\dagger a_{\hexagon 3}^\dagger a_{\hexagon 5}^\dagger a_{\hexagon 6}^{\phantom\dagger} a_{\hexagon 4}^{\phantom\dagger} a_{\hexagon 2}^{\phantom\dagger}+H.c.\big),
\end{eqnarray}
and a term $\hat\mathcal T_k$ can be either a kinetic or a ring-exchange term.
Fig.~\ref{Configuration} shows a possible configuration for the extended partition function associated with this example.

The extended partition function $\mathcal Z(\beta,\tau)$ is a sum of diagonal configurations that belongs to the actual partition
function~$\mathcal Z(\beta)$, and non-diagonal configurations. More precisely,
\begin{eqnarray}
  \nonumber \mathcal Z(\beta,\tau) &=& \mathcal Z(\beta)\\
                                   &+& \!\!\!\sum_{L\neq R}\!\!\textrm{Tr }e^{-(\beta-\tau)\hat\mathcal H}\big|L\big\rangle\big\langle L\big|\hat\mathcal G\big|R\big\rangle\big\langle R\big|e^{-\tau\hat\mathcal H}.
\end{eqnarray}
The purpose of the algorithm is to evolve from a diagonal configuration to another one, via non-diagonal configurations. The role of
the Green operator~$\hat\mathcal G$ is to allow the transition from a configuration to another one by propagating across the operator string
and inserting or removing $\hat\mathcal T$ operators while the offset fluctuates. By satisfying detailed balance~(See section~\ref{GlobalSpaceTimeUpdate}), 
the configurations of the extended partition function~(\ref{ExtendedPartitionFunction4}) can be generated with
an \textit{extended Boltzmann weight}. Then all quantities of interest can be
estimated using those configurations (See section \ref{Measurements}).

\section{\label{GlobalSpaceTimeUpdate}Global space-time update and detailed balance}
\begin{figure}[h]
  \centerline{\includegraphics[width=0.45\textwidth]{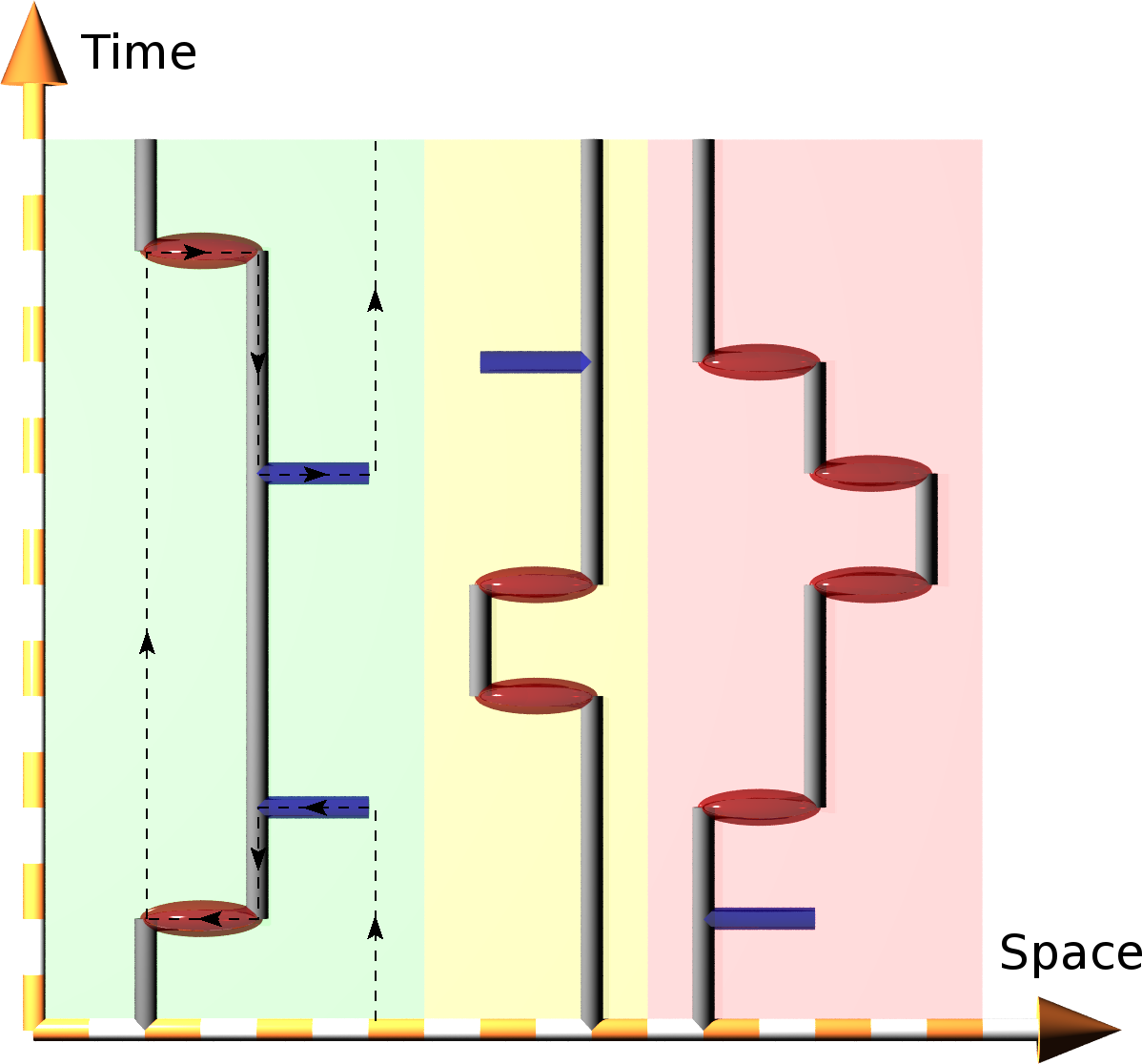}}
  \caption{(Color online) A typical configuration in the SSE representation with a loop update. Diagonal operators (blue)
  do not affect the worldlines (gray), while non-diagonal operators (red) allow to break the worldlines. In order to update the configuration, a loop (black dotted line) is constructed, the occupation numbers of the sites
  that are visited are increased or decreased, and the vertices are updated with diagonal or non-diagonal operators. The loop can jump vertically between two vertices, or horizontally across the same vertex. Because
  horizontal jumps are not permitted between two different vertices, a loop that opens in a given region (cyan, yellow, or pink) must close in the same region. As a result, only a restricted
  region of space can be updated with a single loop. An additional procedure is necessary in order to update the time indices of the vertices,
  and redistribute diagonal operators over the space to give the next loop a chance to update a new region of space.}
  \label{SseUpdate}
\end{figure}

\begin{figure}[h]
  \centerline{\includegraphics[width=0.45\textwidth]{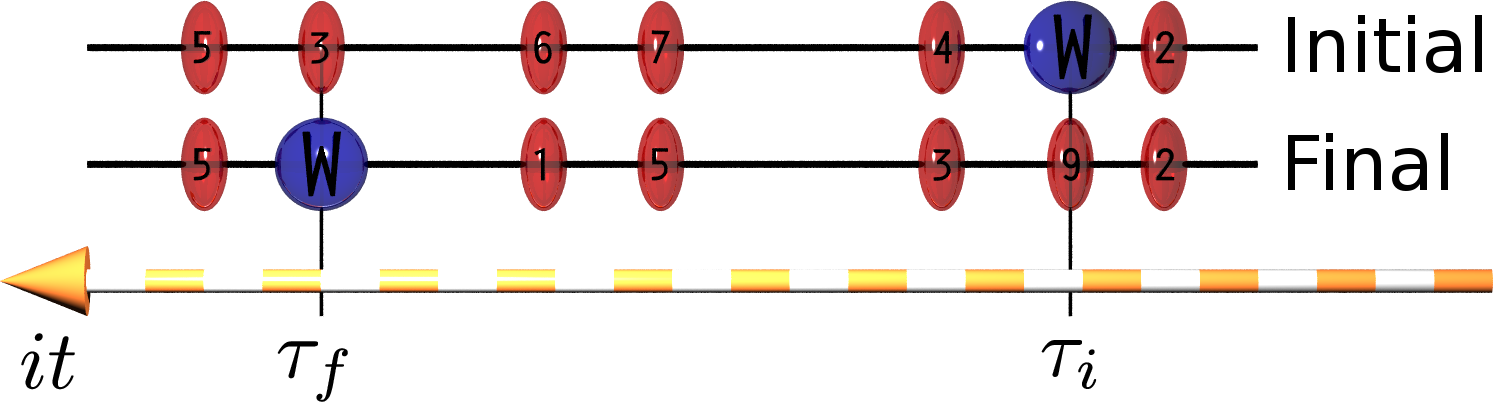}}
  \caption{(Color online) Typical initial and final configurations in the representation of the CW algorithm. Non-diagonal operators (red)
  appear with given space (black labels) and time indices. In order to update the configuration, a worm operator (blue) is propagated in time
  and changes the space indices of the operators that are visited. The worm operator can create an operator with a new time index only at the beginning
  of the propagation, and destroy an operator only at the end. As a result, a single update can affect only a maximum of two time indices, and the
  length of the operator string can fluctuate only by $+1,0,-1$.}
  \label{SpaceUpdate}
\end{figure}

In previous formulations of the SGF algorithm~\cite{SGF,DirectedSGF}, the updating procedure was affecting only space indices
and a maximum of two time indices of the operators of the extended partition function~(\ref{ExtendedPartitionFunction4}).
In this section, we propose a new \textit{directed update} procedure, the ``global space-time update", that is able to globally update both space and time indices of the operators of the
extended partition function. This represents a new improvement, not only over past versions of the SGF algorithm, but also over other algorithms. For example, in the SSE algorithm
a single loop can update only a restricted region of space (Fig.~\ref{SseUpdate}),
and an additional procedure is necessary to update the time indices of the operators. In the CW algorithm, while a single update is able
to affect the entire space, it is unable to update more than two time indices (Fig.~\ref{SpaceUpdate}).

The new global space-time update we propose overcomes those weaknesses by generating a completely new portion of the operator string.
In addition the update is directed, that is to say, the Green operator can be propagated in the same direction over a time window $[\tau_i;\tau_f]$ whose width can be
controlled via an optimization parameter. Also, the length of the operator string can fluctuate by any number with a single update.
Therefore the auto-correlation time between different configurations is reduced, resulting in a better sampling (See section~\ref{Autocorrelation}).

\subsection{Global space-time update scheme}
An easy way to generate configurations of the extended partition function consists in starting from an initial configuration where $\hat\mathcal G$
is the only operator present in the operator string~(\ref{ExtendedPartitionFunction4}), with an arbitrary imaginary time index $\tau$ and
an arbitrary initial state, $\big|L\big\rangle=\big|R\big\rangle$. Then, while propagating to the left (increasing time) or to
the right (decreasing time), the Green operator can either drop a $\hat\mathcal T$ operator behind (creation) or pick up a
$\hat\mathcal T$ operator ahead (destruction). Each creation introduces a new projector $\big|\psi\big\rangle\big\langle\psi\big|$,
and selects a single term of $\hat\mathcal T$. Each destruction removes a projector $\big|\psi\big\rangle\big\langle\psi\big|$
and a single term of $\hat\mathcal T$.
More precisely, assuming a propagation of $\hat\mathcal G$ to the left, a creation corresponds to the following transition
\begin{equation}
  \big\langle L\big|\hat\mathcal G\big|R\big\rangle \rightarrow \big\langle L\big|\hat\mathcal G\big|\psi\big\rangle\big\langle\psi\big|\hat\mathcal T\big|R\big\rangle,
\end{equation}
and a destruction corresponds to
\begin{equation}
  \big\langle L_1\big|\hat\mathcal T\big|L\big\rangle\big\langle L\big|\hat\mathcal G\big|R\big\rangle \rightarrow \big\langle L_1\big|\hat\mathcal G\big|R\big\rangle.
\end{equation}

\begin{figure}[h]
  \centerline{\includegraphics[width=0.45\textwidth]{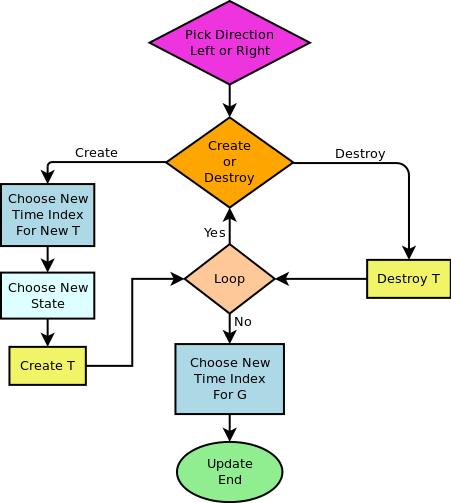}}
  \caption{(Color online) The global space-time update scheme. A $\hat\mathcal T$ operator can be created with any space and time indices,
  and any $\hat\mathcal T$ operator that is encountered can be destroyed. Sequences consisting of several creations and destructions can be
  performed while propagating in the same direction (see text for details).}
  \label{UpdateScheme}
\end{figure}

The idea of the global space-time update is to give to the Green operator a chance to perform several creations and destructions while
propagating in the same direction, and allow both space and time indices of the terms of the $\hat\mathcal T$ operators to be updated.
Thus, while propagating in the same direction, the Green operator can update the system over the whole space in a time window $[\tau_i;\tau_f]$
whose average width can be controlled. The scheme for this global space-time update
is shown in Fig.~\ref{UpdateScheme}. A direction of propagation to the left or to the right is chosen for the Green operator. Then an action, creation or destruction of a $\hat\mathcal T$ operator, is
chosen. If creation is chosen, then a new imaginary time index $\tau^\prime$ is chosen for $\hat\mathcal G$ in the open time window $]\tau_R;\tau_L[$, a state for a new projector is chosen, and a $\hat\mathcal T(\tau^\prime)$
operator is created behind $\hat\mathcal G$. If destruction is chosen, then a $\hat\mathcal T$ operator is destroyed ahead of $\hat\mathcal G$.
After the action has been performed, a decision to loop in the same direction or to stop is made. The algorithm continues to loop
until it chooses to stop. Then a new time index is chosen
in the new open time window $]\tau_R;\tau_L[$ for $\hat\mathcal G$, and the update is over.

Note that when the Green operator crosses the periodic imaginary time boundaries, the state of the trace is updated.
This update ensures the ergodicity of the algorithm, since any permitted term of the $\hat\mathcal T$ operator can be inserted in the operator string at any imaginary
time index. Reciprocally, any term encountered can be removed. Thus any configuration is accessible from a given initial configuration in a finite number of iterations.

\begin{figure}[h]
  \centerline{\includegraphics[width=0.45\textwidth]{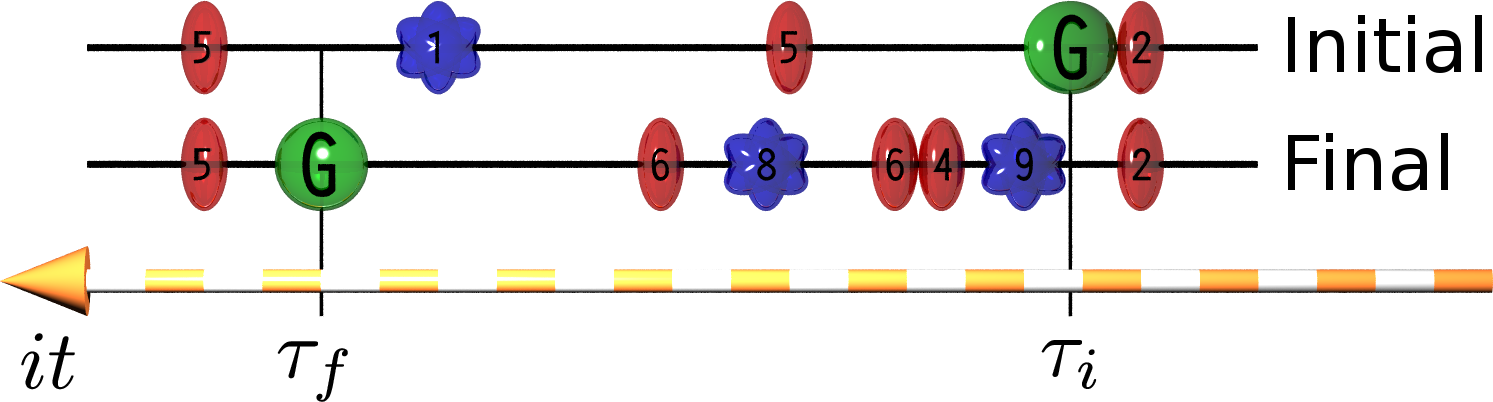}}
  \caption{(Color online) An example of global space-time update for the SGF algorithm in the case of the Hamiltonian~(\ref{HamiltonianExample}). Kinetic (red) and
  ring-exchange (blue) operators can be created or destroyed. The labels on the operators correspond to a particular term (space index).
  The Green operator (green) is propagated to the left from $\tau_i$ to $\tau_f$ and can update both space and time indices of the kinetic and ring-exchange
  operators that are encountered. In addition, the length of the operator string can fluctuate by any number and the average width of the time
  window $[\tau_i;\tau_f]$ is controllable (directed update).}
  \label{SpaceTimeUpdate}
\end{figure}

The advantage of this global space-time update is that it is able to fully update a controllable portion of the operator string.
Fig.~\ref{SpaceTimeUpdate} shows two configurations of the operator string associated with the Hamiltonian~(\ref{HamiltonianExample}) that are accessible from each other with a single update.

\subsection{Detailed balance}
In order to generate operator strings with the (extended) Boltzmann weight, detailed balance must be satisfied.
For this purpose we consider a transition between an initial and a final configuration.

Defining the Boltzmann weight $P_i$ (and $P_f$) of the initial (and final) configuration, and the probability $W_{i\to f}$ (and $W_{f\to i}$) to make a transition from the initial to the final (and from the
final to the initial) configuration, the detailed balance takes the form:
\begin{equation}
  \label{DetailedBalance} P_i W_{i\to f}=P_f W_{f\to i}
\end{equation}
The probabilities $W_{i\to f}$ and $W_{f\to i}$ to make a transition can be factorized as
\begin{eqnarray}
  W_{i\to f} &=& S_{i\to f} A_{i\to f},\\
  W_{f\to i} &=& S_{f\to i} A_{f\to i},
\end{eqnarray}
where $S_{i\to f}$ (resp. $S_{f\to i}$) is the probability to propose the transition from the initial to the final (resp. from the final to the initial) configuration,
and $A_{i\to f}$ (resp. $A_{f\to i}$) is the probability to accept the proposed transition. In the following, we make use of the Metropolis solution
\begin{eqnarray}
  \label{Metropolis} && A_{i\to f}=\min\big(1,q_{i\to f}\big), \\
                     && A_{f\to i}=\min\big(1,q_{f\to i}\big),
\end{eqnarray}
where the acceptance factors $q_{i\to f}$ and $q_{f\to i}$ are given by
\begin{eqnarray}
  \label{AcceptanceFactor} && q_{i\to f}=\frac{P_f S_{f\to i}}{P_i S_{i\to f}}, \\
                           && q_{f\to i}=\frac{P_i S_{i\to f}}{P_f S_{f\to i}}.
\end{eqnarray}

\subsection{Determination of probabilities}
The global space-time update involves several probability functions that need to be determined. The choice of these functions is arbitrary, the only requirements are that ergodicity and detailed balance
must be satisfied. Thus we have some freedom that allows us to make a choice that will be convenient and efficient. We consider the
following probabilities:
\begin{itemize}
  \item $P^{LR}(\sigma)$, with $\sigma=\leftarrow,\rightarrow$, the probability to choose a propagation of the
  Green operator in the $\sigma$ direction, conditioned by the states and time indices with labels $L$ and $R$.
  \item $P^{LR}_\sigma(\dagger)$, the probability to choose a creation, conditioned by $L$, $R$, $\sigma$.
  \item $P^{LR}(\tau)$, the probability to choose a new time index for a $\hat\mathcal T$ operator or for the Green operator, conditioned by $L$, $R$.
  \item $P^{LR}_\sigma(\psi)$, the probability to choose the state $\psi$ for a new projector $\big|\psi\big\rangle\big\langle\psi\big|$, conditioned by $L$, $R$, $\sigma$.
  \item $P^{LR}_\sigma(\circlearrowleft)$, the probability to loop, conditioned by $L$, $R$, $\sigma$.
\end{itemize}
In order to satisfy detailed balance, we must be able to evaluate the acceptance factor $q_\sigma^{\alpha_1\alpha_2\cdots\alpha_n}$
for any sequence consisting of $n$ actions of type $\alpha_i$ in the direction $\sigma$, where $\alpha_i$ is either a creation ($c$) or a destruction ($d$). Our purpose
is to make a suitable choice for the above probabilities, such that all acceptance factors associated with all possible sequences
in any direction reduce to a single acceptance factor $q$.

For this purpose, let us consider a sequence that consists of a propagation to the left with a single creation.
The corresponding acceptance factor is $q_\leftarrow^c$. In the following, we use primed (non-primed) labels for the final (initial)
configuration. The Boltzmann weight $P_i$ of the initial configuration is proportional to
\begin{eqnarray}
  \nonumber               P_i &\propto& \big\langle L\big|\hat\mathcal G(\tau)\big|R\big\rangle\\
  \label{CreationInitial}     &\propto& \big\langle L\big|\hat\mathcal G\big|R\big\rangle e^{\tau(V_L-V_R)}.
\end{eqnarray}
 The hidden matrix elements of the operator string are unchanged by the update. The Boltzmann weight of the final configuration is proportional to
\begin{eqnarray}
  \nonumber P_f &\propto& \big\langle L^\prime\big|\hat\mathcal G(\tau^\prime)\big|R^\prime\big\rangle\big\langle R^\prime\big|\hat\mathcal T(\tau_{R^\prime})\big|R_1^\prime\big\rangle\\
  \nonumber     &\propto& \big\langle L^\prime\big|\hat\mathcal G\big|R^\prime\big\rangle\big\langle R^\prime\big|\hat\mathcal T\big|R_1^\prime\big\rangle\\
                &\times&  e^{\tau^\prime(V_{L^\prime}-V_{R^\prime})} e^{\tau_{R^\prime}(V_{R^\prime}-V_{R_1^\prime})}.
\end{eqnarray}
The probability $S_{i\to f}$ to propose the transition from the initial configuration to the final configuration is the probability to choose a propagation to the left,
times the probability to do a creation, times the probability to choose the time index $\tau_{R^\prime}$ for the new $\hat\mathcal T$, times the probability to choose the new
state $\big|R^\prime\big\rangle$, times the probability to stop the update, times the probability to choose the time index $\tau^\prime$ for $\hat\mathcal G$:
\begin{eqnarray}
  \nonumber S_{i\to f} &=& P^{LR}(\leftarrow)P_\leftarrow^{LR}(\dagger)P^{LR}(\tau_{R^\prime})P_\leftarrow^{LR}(R^\prime)\\
                       &\times& \big(1-P^{L^\prime R^\prime}_\leftarrow(\circlearrowleft)\big)P^{L^\prime R^\prime}(\tau^\prime)
\end{eqnarray}
The probability to propose the reverse update from the final configuration to the initial configuration is simply the probability to choose a propagation to the right,
times the probability to do a destruction, times the probability to stop the update, times the probability to choose the time index $\tau$ for $\hat\mathcal G$:
\begin{eqnarray}
  \nonumber S_{f\to i} &=& P^{L^\prime R^\prime}(\rightarrow)\big(1-P_\rightarrow^{L^\prime R^\prime}(\dagger)\big)\\
                       &\times& \big(1-P^{LR}_\rightarrow(\circlearrowleft)\big)P^{LR}(\tau)
\end{eqnarray}
Putting everything together, and realizing that $V_{L^\prime}=V_L$, and $V_{R_1^\prime}=V_R$, the acceptance factor can be written as:
\begin{eqnarray}
  \nonumber q_\leftarrow^c   &=&      \frac{\big\langle L\big|\hat\mathcal G\big|R^\prime\big\rangle\big\langle R^\prime\big|\hat\mathcal T\big|R\big\rangle}{\big\langle L\big|\hat\mathcal G\big|R\big\rangle P^{LR}_\leftarrow(R^\prime)}\\
  \nonumber                  &\times& \frac{P^{LR}(\tau)}{e^{\tau(V_L-V_R)}} \frac{e^{\tau^\prime(V_{L^\prime}-V_{R^\prime})}}{P^{L^\prime R^\prime}(\tau^\prime)} \frac{e^{\tau_{R^\prime}(V_{R^\prime}-V_{R})}}{P^{LR}(\tau_{R^\prime})}\\
  \label{ExponentiallySmall} &\times& \frac{P^{L^\prime R^\prime}(\rightarrow)\big(1-P^{L^\prime R^\prime}_\rightarrow(\dagger)\big)\big(1-P^{LR}_\rightarrow(\circlearrowleft)\big)}{P^{LR}(\leftarrow)P^{LR}_\leftarrow(\dagger)\big(1-P^{L^\prime R^\prime}_\leftarrow(\circlearrowleft)\big)}
\end{eqnarray}
We notice that the initial and the final times of the Green operator, $\tau$ and $\tau^\prime$, appear only in the second factor of~(\ref{ExponentiallySmall}).
This suggests that it is possible to make the acceptance factor independent of those times by using an exponential distribution
for the time probability,
\begin{equation}
  \label{TimeProbability} P^{LR}(\tau)=\frac{\Delta V e^{\tau\Delta V}}{e^{\tau_L\Delta V}-e^{\tau_R\Delta V}},
\end{equation}
where we have defined $\Delta V=V_L-V_R$ and $\Delta\tau=\tau_L-\tau_R$.

In order to favor the states that are important, we can choose the new states with a probability
that is proportional to the weight of the new matrix elements. Thus we use the following distributions:
\begin{eqnarray}
  \label{StateProbabilityLeft}  P^{LR}_\leftarrow(\psi)  &=& \frac{\big\langle L\big|\hat\mathcal G\big|\psi\big\rangle\big\langle\psi\big|\hat\mathcal T\big|R\big\rangle}{\big\langle L\big|\hat\mathcal G\hat\mathcal T\big|R\big\rangle}\\
  \label{StateProbabilityRight} P^{LR}_\rightarrow(\psi) &=& \frac{\big\langle L\big|\hat\mathcal T\big|\psi\big\rangle\big\langle\psi\big|\hat\mathcal G\big|R\big\rangle}{\big\langle L\big|\hat\mathcal T\hat\mathcal G\big|R\big\rangle}
\end{eqnarray}
Injecting (\ref{TimeProbability}) and (\ref{StateProbabilityLeft}) in (\ref{ExponentiallySmall}), the acceptance factor takes the form
\begin{eqnarray}
  \label{Math1} q_\leftarrow^c &=& \frac{\big\langle L\big|\hat\mathcal G\hat\mathcal T\big|R\big\rangle\big(1-P_\rightarrow^{LR}(\circlearrowleft)\big)}{\big\langle L\big|\hat\mathcal G\big|R\big\rangle P^{LR}(\leftarrow)P_\leftarrow^{LR}(\dagger)}\\
  \nonumber                    &\times& \frac{P^{L^\prime R^\prime}(\rightarrow)\big(1-P^{L^\prime R^\prime}_\rightarrow(\dagger)\big)\big(e^{\Delta\tau^\prime\Delta V^\prime}-1\big)}{\Delta V^\prime\big(1-P^{L^\prime R^\prime}_\leftarrow(\circlearrowleft)\big)},
\end{eqnarray}
and is written as a quantity that depends on the initial configuration, times a quantity that depends on the final configuration. Note that the reverse update corresponds to a propagation
to the right with a destruction, hence $q_\rightarrow^d=1/q_\leftarrow^c$. As a result, the acceptance factor of a left propagation with a destruction, $q_\leftarrow^d$, is obtained by inverting
(\ref{Math1}), switching the direction, and exchanging the primed labels with the non-primed ones:
\begin{eqnarray}
  \nonumber     q_\leftarrow^d &=& \frac{\Delta V\big(1-P^{LR}_\rightarrow(\circlearrowleft)\big)}{P^{LR}(\leftarrow)\big(1-P^{LR}_\leftarrow(\dagger)\big)\big(1-e^{-\Delta\tau\Delta V}\big)}\\
  \label{Math2}                &\times& \frac{\big\langle L^\prime\big|\hat\mathcal G\big|R^\prime\big\rangle P^{L^\prime R^\prime}(\rightarrow)P_\rightarrow^{L^\prime R^\prime}(\dagger)}{\big\langle L^\prime\big|\hat\mathcal T\hat\mathcal G\big|R^\prime\big\rangle\big(1-P_\leftarrow^{L^\prime R^\prime}(\circlearrowleft)\big)}
\end{eqnarray}
For a uniform sampling, we can impose the acceptance factor of a left propagation and creation to be equal to the acceptance factor of a left propagation and destruction, $q_\leftarrow^c=q_\leftarrow^d$.
This  allows us to determine the probability of creation
\begin{eqnarray}
  && P_\leftarrow^{LR}(\dagger)=\frac{\big\langle L\big|\hat\mathcal G\hat\mathcal T\big|R\big\rangle}{\big\langle L\big|\hat\mathcal G\big|R\big\rangle}\frac{1}{f_\leftarrow^{LR}}\\
  && P_\rightarrow^{LR}(\dagger)=\frac{\big\langle L\big|\hat\mathcal T\hat\mathcal G\big|R\big\rangle}{\big\langle L\big|\hat\mathcal G\big|R\big\rangle}\frac{1}{f_\rightarrow^{LR}},
\end{eqnarray}
where we have defined
\begin{eqnarray}
  && f_\leftarrow^{LR}=\frac{\big\langle L\big|\hat\mathcal G\hat\mathcal T\big|R\big\rangle}{\big\langle L\big|\hat\mathcal G\big|R\big\rangle}+\frac{\Delta V}{1-e^{-\Delta\tau\Delta V}}\\
  && f_\rightarrow^{LR}=\frac{\big\langle L\big|\hat\mathcal T\hat\mathcal G\big|R\big\rangle}{\big\langle L\big|\hat\mathcal G\big|R\big\rangle}-\frac{\Delta V}{1-e^{\Delta\tau\Delta V}}.
\end{eqnarray}
The acceptance factors for a creation or a destruction become:
\begin{equation}
  q_\leftarrow^c=q_\leftarrow^d=\frac{\big(1-P^{LR}_\rightarrow(\circlearrowleft)\big)f^{LR}_\leftarrow}{P^{LR}(\leftarrow)}\frac{P^{L^\prime R^\prime}(\rightarrow)}{\big(1-P^{L^\prime R^\prime}_\leftarrow(\circlearrowleft)\big)f^{L^\prime R^\prime}_\rightarrow}
\end{equation}

Let us consider now a propagation to the left and a sequence of two creations, with a corresponding acceptance factor $q_\leftarrow^{cc}$.
The probability of the initial configuration is given by~(\ref{CreationInitial}). The probability of the final configuration is:
\begin{eqnarray}
  \nonumber P_f &\propto& \big\langle L^\prime\big|\hat\mathcal G(\tau^\prime)\big|R^\prime\big\rangle\big\langle R^\prime\big|\hat\mathcal T(\tau_{R^\prime})\big|R_1^\prime\big\rangle\big\langle R_1^\prime\big|\hat\mathcal T(\tau_{R_1^\prime})\big|R_2^\prime\big\rangle\\
  \nonumber     &\propto& \big\langle L^\prime\big|\hat\mathcal G\big|R^\prime\big\rangle\big\langle R^\prime\big|\hat\mathcal T\big|R_1^\prime\big\rangle\big\langle R_1^\prime\big|\hat\mathcal T\big|R_2^\prime\big\rangle\\
                &\times& e^{\tau^\prime(V_{L^\prime}-V_{R^\prime})}e^{\tau_{R^\prime}(V_{R^\prime}-V_{R_1^\prime})}e^{\tau_{R_1^\prime}(V_{R_1^\prime}-V_{R_2^\prime})}
\end{eqnarray}
The probability to propose a transition from the initial configuration to the final configuration is the product:
\begin{eqnarray}
  \nonumber S_{i\to f} &=&      P^{LR}(\leftarrow)P_{\leftarrow}^{LR}(\dagger)P^{LR}(\tau_{R_1^\prime})P^{LR}_\leftarrow(R_1^\prime)\\
  \nonumber            &\times& P^{LR_1^\prime}_\leftarrow(\circlearrowleft)P^{LR_1^\prime}_\leftarrow(\dagger)P^{LR_1^\prime}(\tau_{R^\prime})P^{LR_1^\prime}_\leftarrow(R^\prime)\\
                       &\times& \big(1-P^{L^\prime R^\prime}_\leftarrow(\circlearrowleft)\big)P^{L^\prime R^\prime}(\tau^\prime)
\end{eqnarray}
In the same way, the probability to propose the reverse transition from the final configuration to the initial configuration takes the form:
\begin{eqnarray}
  \nonumber S_{f\to i} &=&      P^{L^\prime R^\prime}(\rightarrow)\big(1-P_\rightarrow^{L^\prime R^\prime}(\dagger)\big)P^{L^\prime R_1^\prime}_\rightarrow(\circlearrowleft)\big(1-P^{L^\prime R_1^\prime}(\dagger)\big)\\
                       &\times& \big(1-P^{LR}_\rightarrow(\circlearrowleft)\big)P^{LR}(\tau)
\end{eqnarray}
Using our previous definitions and the fact that $V_{L^\prime}=V_L$ and $V_{R_2^\prime}=V_R$, the acceptance factor takes the form:
\begin{eqnarray}
  \nonumber q_\leftarrow^{cc} &=&      \frac{\big(1-P^{LR}_\rightarrow(\circlearrowleft)\big)f^{LR}_\leftarrow}{P^{LR}(\leftarrow)}\\
  \nonumber                   &\times& \frac{P^{LR_1^\prime}_\rightarrow(\circlearrowleft)f^{LR_1^\prime}_\leftarrow}{P^{LR_1^\prime}_\leftarrow(\circlearrowleft)f^{LR_1^\prime}_\rightarrow}\\
                              &\times& \frac{P^{L^\prime R^\prime}(\rightarrow)}{\big(1-P^{L^\prime R^\prime}_\leftarrow(\circlearrowleft)\big)f^{L^\prime R^\prime}_\rightarrow}
\end{eqnarray}
and is written as a product of quantities that depend on the initial, intermediate, and final configurations, respectively. One can see from the above expression that a suitable choice for $P_\sigma^{LR}(\circlearrowleft)$
allows us to make the acceptance factor independent of the intermediate configuration. A possible solution is
\begin{equation}
  P_\sigma^{LR}(\circlearrowleft)=\alpha\min\Bigg(1,\frac{f_\sigma^{LR}}{f_{\bar\sigma}^{LR}}\Bigg),
\end{equation}
where $\bar\sigma$ is the opposite direction of $\sigma$, and $\alpha$ is an optimization parameter to be chosen in $[0;1[$. With this choice, the acceptance factor of any update becomes totally independent of the sequence of creations and destructions,
and reads
\begin{equation}
  q_\sigma=\frac{P^{L^\prime R^\prime}(\bar\sigma)Q^{LR}(\sigma)}{P^{LR}(\sigma)Q^{L^\prime R^\prime}(\bar\sigma)},
\end{equation}
with
\begin{equation}
  Q^{LR}(\sigma)=f^{LR}_\sigma\Big(1-\alpha\min\big(1,f^{LR}_{\bar\sigma}/f^{LR}_\sigma\big)\Big).
\end{equation}
Finally we can impose the acceptance factor of a propagation to the left to be equal to the acceptance factor of a propagation to the right, $q_\leftarrow=q_\rightarrow$. This is realized if
\begin{equation}
  P^{LR}(\sigma)=\frac{Q^{LR}(\sigma)}{Q^{LR}(\sigma)+Q^{LR}(\bar\sigma)},
\end{equation}
and, defining $Q^{LR}=Q^{LR}(\leftarrow)+Q^{LR}(\rightarrow)$, we are left with a single acceptance factor for any update:
\begin{equation}
  \label{Math4} q=\frac{Q^{LR}}{Q^{L^\prime R^\prime}}
\end{equation}
Because the acceptance factor (\ref{Math4}) is written as a ratio of a quantity that depends only on the initial configuration and a quantity that depends only on the final configuration, an ultimate
simplification can be done. Using (\ref{Math4}) and defining $Q_i=Q^{LR}$ and $Q_f=Q^{L^\prime R^\prime}$, we can rewrite (\ref{AcceptanceFactor}) as:
\begin{equation}
  \label{AcceptanceFactor2} \frac{P_fQ_fS_{f\to i}}{P_iQ_iS_{i\to f}}=1
\end{equation}
The above equation can be interpreted as follows: Accepting all transitions with a probability of 1 is equivalent to sampling the partition function with the pseudo-Boltzmann weight $PQ$ instead
of the true Boltzmann weight $P$. The statistical average of any operator $\hat\mathcal A$ can be obtained with a simple renormalization:
\begin{equation}
  \label{Renormalization} \big\langle\hat\mathcal A\big\rangle_P=\frac{\big\langle\hat\mathcal A/Q\big\rangle_{PQ}}{\big\langle 1/Q\big\rangle_{PQ}}
\end{equation}
By construction, the function $Q$ never diverges nor vanishes. Hence the renormalization is well defined in any case.
Note that we have explicitly excluded the value $1$ from the allowed
values for $\alpha$. This prevents the probability of looping, $P_\sigma^{LR}(\circlearrowleft)$, from being systematically equal to the unity
in diagonal configurations, otherwise no measurements would be possible. 
The advantage of accepting all transitions with a probability of $1$ is that no CPU time is wasted with useless rejected updates, and there is no need to record the changes made in the operator string during the update.
Also, accepting a ``bad" transition from time to time may help the system to escape from a local minimum of the energy.

It is worth to emphasize the importance of interpreting correctly the meaning of $\Delta\tau=\tau_L-\tau_R$, especially at high
temperature where the system is dominated by configurations with zero or a single $\hat\mathcal T$ operator. $\Delta\tau$ corresponds
to the time length over which the Green operator can be shifted without encountering a $\hat\mathcal T$ operator. On the one hand, when there is a single $\hat\mathcal T$ operator in the string, then
$\tau_L=\tau_R$ and the Green operator is able to move everywhere over the imaginary time axis, except over the single point
$\tau_L=\tau_R$. Thus the difference $\tau_L-\tau_R$ must include the periodicity of the imaginary time axis, leading to the
value $\Delta\tau=\beta$. On the other hand, when the operator string is empty, there is no limit of the time length over which the
Green operator can be shifted. So the value $\Delta\tau=+\infty$ must be used, which ensures that the probability of creation of
a $\hat\mathcal T$ operator is $P_\sigma^{LR}(\dagger)=1$. In this limit, the probability distribution for the time index of the
new $\hat\mathcal T$ operator or the Green operator becomes defined in the interval $]-\infty;+\infty[$ with the value $P^{LR}(\tau)\to 0$.
Taking into account the periodicity of the imaginary time axis allows us to restrict the distribution over the finite range $[0;\beta[$
with $P^{LR}(\tau)=1/\beta$. However it is crucial to use $\Delta\tau=+\infty$ in all other equations where $\Delta\tau$ is involved,
in order to have them correct.

By adjusting the value of $\alpha$ for the probability to loop, one can tune the ``directionality" of the update, that is to say the average length of the sequence of creations
and destructions of the update. Having long sequences of creations and destructions reduces the probability for an update to undo the changes
made in the configuration with the previous update. This also increases the width of the time window $[\tau_i,\tau_f]$ of the operator string that is updated, and 
reduces the auto-correlation time (see section~\ref{Autocorrelation}).

The choice of the function $g(n)$ determines how the extended space of configurations is sampled. In section~\ref{Measurements}, it
is shown that it is necessary to generate diagonal configurations in order to make measurements. Thus $g(n)$ must be a function that
decreases with $n$ sufficiently fast in order to have a chance to generate diagonal configurations. However, non-diagonal
configurations are needed in order to update the system. The choice of $g(n)$ must be done is such a way that all non-diagonal terms
have a comparable probability to be introduced in the operator string. In practice, we find that the choice $g(n)=1/L^{n/2}$ where
$L$ is the number of lattice sites is a
good choice for Hamiltonians for which the highest order of non-diagonal terms is 2, like the Hamiltonian~(\ref{BoseHubbard}).
For our non-trivial example~(\ref{HamiltonianExample}), a better choice is:
\begin{equation}
  g(n)=
  \left\lbrace
  \begin{array}{ll}
    1 & n=0\\
    1/L & n=1,2,3,4\\
    1/L^{n-3} & n\geq 5
  \end{array}
  \right.
\end{equation}

\subsection{Summary of probabilities}
\label{summary}
The total weight of creation and destruction for a particular direction of motion,
$\sigma=\rightarrow,\leftarrow$, 
and for particular $\big|L\big\rangle$ and $\big|R\big\rangle$ states, can be parameterized as:
\begin{equation}
W_{\sigma}=W_{\sigma}^{+}+W_{\sigma}^{-},
\end{equation}
where $W_{\sigma}^{+}$ is the weight of creation, 
\begin{eqnarray}
W_{\leftarrow}^{+} & = & \frac{\big\langle L\big|\hat\mathcal G\hat\mathcal T\big|R\big\rangle}{\big\langle L\big|\hat\mathcal G\big|R\big\rangle}\\
W_{\rightarrow}^{+} & = & \frac{\big\langle L\big|\hat\mathcal T\hat\mathcal G\big|R\big\rangle}{\big\langle L\big|\hat\mathcal G\big|R\big\rangle}
\end{eqnarray}
and $W_{\sigma}^{-}$ is the weight of destruction, 
\begin{equation}
W_{\sigma}^{-}=\frac{s_\sigma\Delta V}{1-e^{s_\sigma\Delta\tau\Delta V}},
\end{equation}
where we introduced the symbol $s_{\sigma}$ with $s_{\leftarrow}=-1$ and $s_{\rightarrow}=+1$.

\begin{figure}[h]
\centerline{\includegraphics[width=0.45\textwidth]{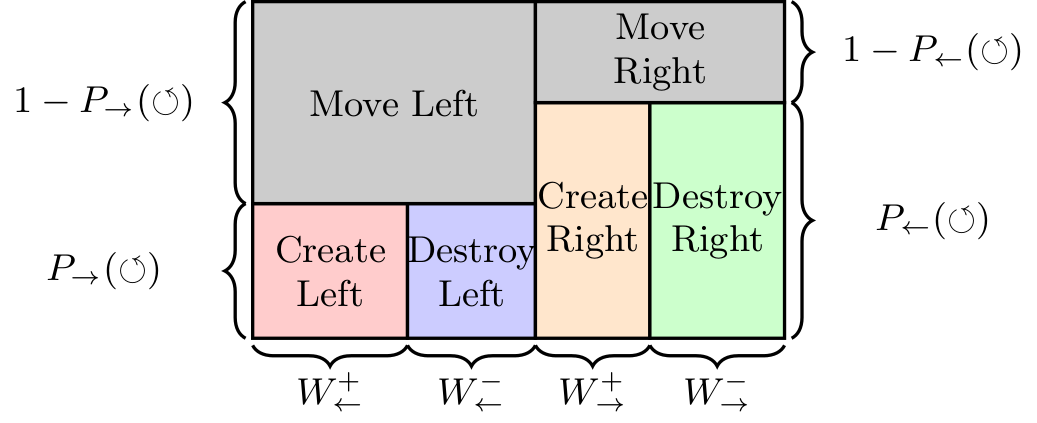}}
\caption{(color online) Graphical representation of the 
update probabilities. First the algorithm picks a direction 
of motion with a relative probability represented
by the gray tiles. Their total area is the renormalization weight. Once
the direction is fixed it will pick a destruction or creation of an
operator with relative probabilities represented by the colored
tiles of the corresponding side. The height of those tiles represents
the probability to keep moving in the same direction after a creation
or destruction.\label{fig:probabilitymap}}
\end{figure}

In this notation the absolute creation probability is
\begin{eqnarray}
P_{\sigma}\left(+\right) & = & \frac{W_{\sigma}^{+}}{W_{\sigma}}
\end{eqnarray}
Equivalently the destruction probability is
\begin{equation}
P_{\sigma}\left(-\right)=\frac{W_{\sigma}^{-}}{W_{\sigma}}
\end{equation}
The probability to loop is
\begin{eqnarray}
P_{\sigma}\left(\circlearrowleft\right) & = & \alpha\min\left(1,\frac{W_{\sigma}}{W_{\overline{\sigma}}}\right).
\end{eqnarray}
The \emph{relative} probability to pick a particular direction reads 
\begin{equation}
Q\left(\sigma\right)=W_{\sigma}\big(1-P_{\overline{\sigma}}(\circlearrowleft)\big).
\end{equation}
As shown in Fig.~\ref{UpdateScheme}, the algorithm starts by picking 
a particular direction with relative probability $Q(\sigma)$. 
Then it decides if it will destroy or create an operator with 
relative probabilities $W^{+}_{\sigma}$ and $W^{-}_{\sigma}$ respectively. If creation
is chosen the new time index is chosen within the open interval $]\tau_R,\tau_L[$
with relative probability distribution $e^{\tau\Delta V}$ and the space index
with relative probability given by Eq.~(\ref{StateProbabilityLeft}) and~(\ref{StateProbabilityRight}).
At the end of the creation or destruction the algorithm decides if it wants to continue, 
with absolute probability  $P_{\sigma}\left(\circlearrowleft\right)$, or stop and start over.
All the update probabilities are summarized in table \ref{tab:summary_probabilities}
and shown graphically in Fig. \ref{fig:probabilitymap}.
\begin{table}
\centering{}%
\begin{tabular}{|l|l|}
\hline 
Destruction weight & $W_{\sigma}^{-}=\frac{s_\sigma\Delta V}{1-e^{s_\sigma\Delta\tau\Delta V}}$\tabularnewline
\hline 
Creation weight & \parbox{4cm}{$W_{\leftarrow}^{+}  =  \frac{\big\langle L\big|\hat\mathcal G\hat\mathcal T\big|R\big\rangle}{\big\langle L\big|\hat\mathcal G\big|R\big\rangle}$\\
 $W_{\rightarrow}^{+}  =  \frac{\big\langle L\big|\hat\mathcal T\hat\mathcal G\big|R\big\rangle}{\big\langle L\big|\hat\mathcal G\big|R\big\rangle}$}\tabularnewline
\hline 
Loop probability& $P_{\sigma}\left(\circlearrowleft\right)=\alpha\min\left(1,\frac{W_{\sigma}^{-}+W_{\sigma}^{+}}{W_{\overline{\sigma}}^{-}+W_{\overline{\sigma}}^{+}}\right)$\tabularnewline
\hline 
Direction weight & $Q\left(\sigma\right)=\left(W_{\sigma}^{-}+W_{\sigma}^{+}\right)\left(1-P_{\overline{\sigma}}\left(\circlearrowleft\right)\right)$\tabularnewline
\hline 
Renormalization weight & $Q=Q\left(\leftarrow\right)+Q\left(\rightarrow\right)$\tabularnewline
\hline 
Time index weight & $P\left(\tau\right)=\frac{\Delta V e^{\tau\Delta V}}{e^{\tau_L\Delta V}-e^{\tau_R\Delta V}}$\tabularnewline
\hline 
Space index weight & 
\parbox{4cm}{
  $P^{LR}_\leftarrow(\psi)  = \frac{\big\langle L\big|\hat\mathcal G\big|\psi\big\rangle\big\langle\psi\big|\hat\mathcal T\big|R\big\rangle}{\big\langle L\big|\hat\mathcal G\hat\mathcal T\big|R\big\rangle}$\\
  $P^{LR}_\rightarrow(\psi) = \frac{\big\langle L\big|\hat\mathcal T\big|\psi\big\rangle\big\langle\psi\big|\hat\mathcal G\big|R\big\rangle}{\big\langle L\big|\hat\mathcal T\hat\mathcal G\big|R\big\rangle}$
}

\tabularnewline
\hline 
\end{tabular}\caption{Summary of the update probabilities.\label{tab:summary_probabilities}}
\end{table}

\section{\label{Measurements} Measurements}
\subsection{Notation}
We will systematically use the notations $\tilde\mathcal A(\tau)$ and $\hat\mathcal A(\tau)$ for the Heisenberg representation and
the interaction representation of an operator $\hat\mathcal A$, respectively:
\begin{eqnarray}
  && \tilde\mathcal A(\tau)=e^{\tau\hat\mathcal H}\hat\mathcal A e^{-\tau\hat\mathcal H}\\
  && \hat\mathcal A(\tau)=e^{\tau\hat\mathcal V}\hat\mathcal A e^{-\tau\hat\mathcal V}
\end{eqnarray}
It is important to avoid confusion between different types of statistical averages. We use the two following
notations and definitions:
\begin{eqnarray}
  \label{HeisenbergRepresentation}  && \big\langle\hat\mathcal A\big\rangle^h=\frac{1}{\mathcal Z}\textrm{Tr }\hat\mathcal A e^{-\beta\hat\mathcal H}\\
  \label{InteractionRepresentation} && \big\langle\hat\mathcal A\big\rangle^i=\frac{1}{\mathcal Z}\textrm{Tr }e^{-\beta\hat\mathcal V}\stackrel{\leftarrow}{\textrm{\bf T}_\tau}\Big[\hat\mathcal A e^{\int_0^\beta\hat\mathcal T(\tau)d\tau}\Big]
\end{eqnarray}
These definitions ensure that
\begin{equation}
  \big\langle\tilde\mathcal A(\tau)\big\rangle^h=\big\langle\hat\mathcal A(\tau)\big\rangle^i,
\end{equation}
and in particular, for an operator $\hat\mathcal A$ independent of imaginary time, the two statistical averages
are equivalent and the superscript can be omitted,
\begin{equation}
  \big\langle\hat\mathcal A\big\rangle=\big\langle\hat\mathcal A\big\rangle^h=\big\langle\hat\mathcal A\big\rangle^i.
\end{equation}
We show in appendix that, for any product of operators $\hat\mathcal A_1,\cdots,\hat\mathcal A_R$ and any set of time indices
$\tau_1\cdots,\tau_R$, we have:
\begin{equation}
  \label{HeisenbergInteraction} \big\langle\stackrel{\leftarrow}{\textrm{\bf T}_\tau}\tilde\mathcal A_R(\tau_R)\cdots\tilde\mathcal A_1(\tau_1)\big\rangle^h=\big\langle\hat\mathcal A_R(\tau_R)\cdots\hat\mathcal A_1(\tau_1)\big\rangle^i
\end{equation}

\subsection{Measuring quantities with the extended Boltzmann weight}
By definition of the extended partition function~(\ref{ExtendedPartitionFunction1}), the states $\big|L\big\rangle$ and
$\big|R\big\rangle$ of the Green operator are associated with the \textit{extended} Boltzmann weight:
\begin{equation}
  P(L,R)=\frac{\big\langle L\big|\hat\mathcal G\big|R\big\rangle\big\langle R\big|e^{-\beta\hat\mathcal H}\big|L\big\rangle}{\mathcal Z(\beta,\tau)}
\end{equation}
This extended Boltzmann weight can be used to measure any operator $\hat\mathcal A$. Indeed, consider the expectation value of
$\hat\mathcal A$:
\begin{eqnarray}
  \nonumber \big\langle\hat\mathcal A\big\rangle &=& \frac{\textrm{Tr }\hat\mathcal A e^{-\beta\hat\mathcal H}}{\mathcal Z(\beta)}\\
  \label{Expectation}                            &=& \frac{\sum_{\psi}\big\langle\psi\big|\hat\mathcal A e^{-\beta\hat\mathcal H}\big|\psi\big\rangle}{\mathcal Z(\beta)}
\end{eqnarray}
By introducing a complete set of states, and using the fact that all diagonal matrix elements of $\hat\mathcal G$ are equal to~$1$, Eq.~(\ref{Expectation}) can be rewritten as
\begin{eqnarray}
  \nonumber \big\langle\hat\mathcal A\big\rangle &=& \frac{\sum\limits_{L,R}\big\langle L\big|\hat\mathcal A\big|R\big\rangle\big\langle R\big|e^{-\beta\hat\mathcal H}\big|L\big\rangle}{\mathcal Z(\beta)}\\
                                                 &=& \frac{\sum\limits_{L,R}\frac{\big\langle L\big|\hat\mathcal A\big|R\big\rangle}{\big\langle L\big|\hat\mathcal G\big|R\big\rangle}P(L,R)}{\sum\limits_{L,R}\delta_{L,R}P(L,R)},
\end{eqnarray}
where $\delta_{L,R}$ is the Kr\"onecker delta. By performing an importance sampling (denoted $S_{LR}$) over $S$ samples of states $\big|L\big\rangle$ and
$\big|R\big\rangle$ with the distribution $P(L,R)$, the expectation value reduces to
\begin{eqnarray}
  \nonumber \big\langle\hat\mathcal A\big\rangle &=& \lim_{S\to\infty}\frac{\sum\limits_{S_{LR}}\frac{\big\langle L\big|\hat\mathcal A\big|R\big\rangle}{\big\langle L\big|\hat\mathcal G\big|R\big\rangle}}{\sum\limits_{S_{LR}}\delta_{L,R}}\\
                                                 &=& \lim_{S\to\infty}\frac{1}{N_d}\sum\limits_{S_{LR}}\frac{\big\langle L\big|\hat\mathcal A\big|R\big\rangle}{\big\langle L\big|\hat\mathcal G\big|R\big\rangle},
\end{eqnarray}
where $N_d$ is the number of diagonal configurations in the set of samples.

If we make the choice of accepting all updates with a probability of $100\%$, it is necessary to perform the renormalization~(\ref{Renormalization}).
The estimator for $\hat\mathcal A$ becomes:
\begin{equation}
  \label{ExpectationValue}\big\langle\hat\mathcal A\big\rangle=\lim_{S\to\infty}\frac{\sum\limits_{S_{LR}}\frac{\big\langle L\big|\hat\mathcal A/Q\big|R\big\rangle}{\big\langle L\big|\hat\mathcal G\big|R\big\rangle}}{\sum\limits_{S_\psi}\big\langle\psi\big|1/Q\big|\psi\big\rangle}
\end{equation}
It is important to note that measurements can be done only at the end of a global space-time update, when the loop is over, that is
to say when detailed balance is satisfied.

\subsection{Quantities represented by diagonal operators}
For diagonal operators, only diagonal configurations $\big|L\big\rangle=\big|R\big\rangle$ have a non-vanishing contribution,
and Eq.(\ref{ExpectationValue}) reduces to:
\begin{equation}
  \label{MeasDiagonal}\big\langle\hat\mathcal A\big\rangle=\lim_{S\to\infty}\frac{\sum\limits_{S_\psi}\big\langle\psi\big|\hat\mathcal A/Q\big|\psi\big\rangle}{\sum\limits_{S_\psi}\big\langle\psi\big|1/Q\big|\psi\big\rangle}
\end{equation}
This is the case for the diagonal energy $\hat\mathcal V$~(\ref{MeasEnergy}), the superfluid density~(\ref{MeasSuperfluid} and \ref{ImprovedSuperfluid}), imaginary time-dependent
density-density correlation functions~(\ref{MeasDenDenCor}), and the imaginary dynamical structure factor~(\ref{MeasDynamical}).

\subsection{Quantities represented by non-diagonal operators}
Non-diagonal operators can be measured ``on the fly" with Eq.(\ref{ExpectationValue}), while exploring the extended space of
configurations. The numerator has contributions from
non-diagonal configurations $\big|L\big\rangle\neq\big|R\big\rangle$, while the denominator is evaluated in diagonal
configurations $\big|L\big\rangle=\big|R\big\rangle$. This is the case for Green functions and the momentum distribution function~(\ref{MeasGreen}). The
non-diagonal energy $\hat\mathcal T$ can also be measured by decomposing it as a sum of Green functions. However it is possible to
measure it using diagonal configurations only~(\ref{MeasEnergy}), which is simpler and more efficient.

\subsection{Integrals over imaginary time}
\subsubsection{Diagonal operators}
Many quantities of interest are defined as integrals of the form
\begin{equation}
  \label{Integral} \hat\mathcal I=\big\langle\int_0^\beta\hat\mathcal O_d(\tau)f(\tau)d\tau\big\rangle^i,
\end{equation}
where $\hat\mathcal O_d$ is a diagonal operator and $f(\tau)$ an arbitrary function. This is the case, for instance, for the
improved estimator of the diagonal energy (see~\ref{MeasEnergy}) and the Fourier transform of the local density
(see~\ref{MeasDynamical}). Therefore, it is necessary to understand how integrals like~(\ref{Integral}) can be evaluated exactly
within the SGF framework.

For this purpose, consider a configuration $\mathcal C_n$ of the operator string with length $n$, imaginary time indices $\tau_1,\tau_2,\cdots,\tau_n$, and the convention
that $\tau_0=0$ and $\tau_{n+1}=\beta$. For any time index $\tau\in\big[0;\beta\big[$, there exists a unique $k$ such that $\tau_k\leq\tau<\tau_{k+1}$.
This implies the identity
\begin{equation}
  \label{Identity} \sum_{k=0}^n\Theta(\tau_k\leq\tau<\tau_{k+1})=1\quad\forall\tau\in\big[0;\beta\big[,
\end{equation}
where $\Theta(arg)=1$ if $arg$ is true, $0$ otherwise. We denote the value of the integral~({\ref{Integral}) in this particular configuration
by:
\begin{equation}
  \nonumber\big\langle\int_0^\beta\hat\mathcal O_d(\tau)f(\tau)d\tau\big\rangle_{\mathcal C_n}^i
\end{equation}
This integral receives only contributions from diagonal configurations, where all worldlines are straight between two consecutive
time indices, $\tau_k$ and $\tau_{k+1}$. As a result, $\hat\mathcal O_d$ has a constant expectation value between $\tau_k$ and $\tau_{k+1}$ and, using~(\ref{Identity}),
the integral of $\hat\mathcal O_d$ in the configuration $\mathcal C_n$ takes the form
\begin{eqnarray}
  \nonumber\big\langle\int_0^\beta\hat\mathcal O_d(\tau)f(\tau)d\tau\big\rangle_{\mathcal C_n}^i &=&      \sum_{k=0}^n\big\langle\psi_k^{k+1}\big|\hat\mathcal O_d\big|\psi_k^{k+1}\big\rangle\\
  \label{DiagonalPart}                                                                                  &\times& \int_{\tau_k}^{\tau_{k+1}}\!\!\!\!\!\!\!\!f(\tau)d\tau,
\end{eqnarray}
where $\psi_k^{k+1}$ labels the state between the time indices $\tau_k$ and $\tau_{k+1}$.

\subsubsection{Non-diagonal operators}
The integral over the imaginary time axis of a non diagonal operator $\hat\mathcal O_{nd}$ can be easily evaluated if $\hat\mathcal O_{nd}$
is a particular Hamiltonian term, $\hat\mathcal O_{nd}=\hat\mathcal T_k$. In this case, for a particular configuration $\mathcal C_n$, we have
\begin{equation}
  \nonumber\big\langle\int_0^\beta\hat\mathcal T_k(\tau)d\tau\big\rangle_{\mathcal C_n}^i=\frac{n_k}{\beta},
\end{equation}
where $n_k$ is the number of ocurences of $\hat\mathcal T_k$ in the operator string. This is a particular
case of the theorem presented in next paragraph.

\subsubsection{Theorem of integration}
We propose here a convenient theorem that allows us to easily measure time integrals of multi-point correlation functions.
We consider a set of $M$ arbitrary diagonal operators $\hat\mathcal D_k$, and a set of $N$ distinct non-diagonal Hamiltonian terms $\hat\mathcal T_k$
each being associated with an integer $p_k$. For a given operator $\hat\mathcal O$, we define the integrals:
\begin{eqnarray}
  && \tilde\mathcal I(\hat\mathcal O)=\frac{1}{\beta}\int_0^\beta \tilde\mathcal O(\tau)d\tau\\
  && \hat\mathcal I(\hat\mathcal O)=\frac{1}{\beta}\int_0^\beta \hat\mathcal O(\tau)d\tau
\end{eqnarray}
Then we have
\begin{eqnarray}
  \label{TheoremEquation}\nonumber &&\Big\langle\stackrel{\leftarrow}{\textrm{\bf T}_\tau}\Big[\tilde\mathcal I(\hat\mathcal D_M)\cdots\tilde\mathcal I(\hat\mathcal D_1)\big(\tilde\mathcal I(\hat\mathcal T_N)\big)^{p_N}\cdots\big(\tilde\mathcal I(\hat\mathcal T_1)\big)^{p_1}\Big]\Big\rangle^h\\
  &&\quad\quad=\frac{\Big\langle\hat\mathcal I(\hat\mathcal D_M)\cdots\hat\mathcal I(\hat\mathcal D_1)\frac{n_N!}{(n_N-p_N)!}\cdots\frac{n_1!}{(n_1-p_1)!}\Big\rangle^i}{\beta^{p_N+\cdots+p_1}},
\end{eqnarray}
where $n_k$ is the number of $\hat\mathcal T_k$ terms in the configuration. Each integral $\hat\mathcal I(\hat\mathcal D_k)$ in the
RHS of~(\ref{TheoremEquation}) is given by~(\ref{DiagonalPart}).

This theorem states that it is equivalent to measure a time-ordered product of operators in the Heisenberg representation
and to measure the same product of operators in the interaction representation. The simplification provided by the theorem is that the
measurement of the time integral of a non-diagonal Hamiltonian term reduces to counting the number of occurences of that term
in the operator string. More precisely, the factor $\frac{n_k!}{(n_k-p_k)!}$ corresponds the number of distinct possibilities to
select $p_k$ operators of type $\hat\mathcal T_k$ in the operator string. The exponent of $\beta$ in the denominator is the total number
of integrals of non-diagonal Hamiltonian terms.

The proof of this theorem is given in appendix, and
a direct application of it allows us to measure the energy of the system
(see~\ref{MeasEnergy}) and the specific-heat (see~\ref{MeasSpecific}).

\subsection{Particular measurements}
\subsubsection{\label{MeasEnergy}The energy}
The diagonal energy $\hat\mathcal V$ can be measured directly in diagonal configurations by using~(\ref{MeasDiagonal}).
However, a better estimator can be constructed by taking advantage of the invariance by imaginary time translation, and integrating over
the full imaginary time axis.
Using the theorem (\ref{TheoremEquation}) with $M=1$, $N=0$, and $\hat\mathcal D=\hat\mathcal V$, we have:
\begin{eqnarray}
  \nonumber\big\langle\hat\mathcal V\big\rangle &=& \frac{1}{\beta}\big\langle\stackrel{\leftarrow}{\textrm{\bf T}_\tau}\int_0^\beta\tilde\mathcal V(\tau)d\tau\big\rangle^h\\
  \label{ImprovedDiagonal}                      &=& \frac{1}{\beta}\big\langle\int_0^\beta\hat\mathcal V(\tau)d\tau\big\rangle^i
\end{eqnarray}
The integral in (\ref{ImprovedDiagonal}) can be evaluated using (\ref{DiagonalPart}) with $\hat\mathcal O_d=\hat\mathcal V$ and $f(\tau)=1$.

The non-diagonal energy $\hat\mathcal T$ can be measured easily by averaging the length of the operator string. More generally, the energy associated
with any non-diagonal Hamiltonian term can be measured by counting the number of occurrences of that term in the operator string with
diagonal configurations. Indeed, consider the decomposition:
\begin{equation}
  \hat\mathcal T=\sum_k \hat\mathcal T_k
\end{equation}
Using (\ref{TheoremEquation}) with $M=0$ and $N=1$, the expectation value of a particular term $\hat\mathcal T_k$ is given by:
\begin{eqnarray}
  \nonumber \big\langle\hat\mathcal T_k\big\rangle &=& \frac{1}{\beta}\big\langle\stackrel{\leftarrow}{\textrm{\bf T}_\tau}\int_0^\beta\tilde\mathcal T_k(\tau)d\tau\big\rangle^h\\
  \label{Labelwithnoname}                          &=& \frac{1}{\beta}\big\langle n_k\big\rangle^i
\end{eqnarray}
In particular, the non-diagonal energy $\hat\mathcal T$ is given by averaging the total length $n$ of the operator string:
\begin{equation}
  \big\langle\hat\mathcal T\big\rangle=\frac{1}{\beta}\big\langle n\big\rangle^i
\end{equation}

\subsubsection{\label{MeasSuperfluid}The superfluid density at finite temperature}
The superfluid density can be easily measured via the winding number. For a $d$-dimensional system, the winding number is a vectorial
operator, $\hat W$, whose components measure the number of times that the worldlines cross the boundaries of the system in
the primary directions of the lattice.
Considering a configuration $\mathcal C_n$ of the operator string with length $n$ and imaginary time indices $\tau_1,\tau_2,\cdots,\tau_n$, we define the
associated \textit{pseudo-current} in the $\zeta$ direction at time $\tau$ as
\begin{equation}
  \label{PseudoCurrent} \hat j_\zeta(\tau)=\sum_{k=1}^n \Delta_k^\zeta\delta(\tau-\tau_k),
\end{equation}
where $\Delta_k^\zeta$ measures the discontinuity of the worldlines in the $\zeta$ direction introduced by the $\hat\mathcal T$ operator acting at
time $\tau_k$. In our example~(\ref{HamiltonianExample}),
a kinetic term acting in the $\zeta$ direction gives $\Delta_k^\zeta=\pm 1$, while a ring-exchange term
systematically gives $\Delta_k^\zeta=0$. By definition, the winding in the $\zeta$ direction is obtained by integrating the pseudo-current
\begin{eqnarray}
  \nonumber\hat W_\zeta &=& \frac{1}{L_\zeta}\int_0^\beta \hat j_\zeta(\tau)d\tau\\
  \label{WindingPseudoCurrent} &=& \frac{1}{L_\zeta}\sum_{k=1}^n\Delta_k^\zeta,
\end{eqnarray}
where $L_\zeta$ is the number of sites in the $\zeta$ direction. Thus the winding can be easily measured in the configuration
$\mathcal C_n$ by using Eq.(\ref{WindingPseudoCurrent}).
For identical particles in a cubic lattice with $L^d$ sites, it has been shown~\cite{Pollock} that the
dimensionless superfluid density is given by
\begin{equation}
  \label{FiniteTemperatureSuperfluid} \rho_s=\frac{L^{2-d}\big\langle\hat W^2\big\rangle}{2dt\beta},
\end{equation}
where $t=\frac{\hbar^2}{2ma^2}$, $m$ is the mass of one particle, and $a$ is the lattice constant.

\subsubsection{\label{ImprovedSuperfluid}Improved estimator for the zero-temperature superfluid density}
As we have seen above, the superfluid density $\rho_s$ can be measured via the winding number $\hat W$. However $\rho_s$ can
show a strong dependence in temperature, which makes it difficult to estimate its zero-temperature value. Measuring the zero-temperature
superfluid density requires, in principle, to perform simulations with increasing values of the inverse temperature $\beta$, which is computationally expensive,
and then perform an extrapolation to $\beta\to\infty$. We propose here an improved estimator that has a faster convergence to
the zero-temperature superfluid density, thus making simulations easier.
This improved estimator has been proposed by Batrouni and Scalettar for the discrete time Worldline algorithm~\cite{Batrouni}. The
generalization to continuous time is straightforward using our continuous-time definition of the pseudo current~(\ref{PseudoCurrent}).
The improved estimator is actually for the winding number, and we determine the
superfluid density using~(\ref{FiniteTemperatureSuperfluid}).

To this end, we define the Fourier transform of the pseudo-current
\begin{eqnarray}
  \nonumber\mathring j_\zeta(\omega_p) &=& \int_0^\beta\hat j_\zeta(\tau)e^{-i\omega_p\tau}d\tau\\
                                     &=& \sum_{k=1}^n \Delta_k^\zeta e^{-i\omega_p\tau_k},
\end{eqnarray}
with $\omega_p=\frac{2\pi p}{\beta}$. One notices that $\hat W_\zeta=\mathring j_\zeta(0)/L_\zeta$.
The idea of the improved estimator $\hat W_\zeta^{i}$ is based on the fact that the discrete frequencies $\omega_p$ become
continuous in the limit $\beta\to\infty$, and the ``numerical observation" that the values of $\big|\mathring j_\zeta(\omega_p)\big|^2$ for
$p\geq 1$ scale linearly with $\beta$ at finite but low temperature. As a result, the value of $\big|\mathring j_\zeta(0)\big|^2$ at zero-temperature
can be estimated at finite temperature by performing a linear extrapolation.

Therefore we define the improved estimator for the zero-temperature winding number as:
\begin{equation}
  \label{ImprovedWinding} W_\zeta^{i\:2}=\frac{1}{L_\zeta^2}\Big(2\big|\mathring j_\zeta(\omega_1)\big|^2-\big|\mathring j_\zeta(\omega_2)\big|^2\Big)
\end{equation}

Since both $\omega_1$ and $\omega_2$ vanish in the limit $\beta\to\infty$, we have $\lim_{\beta\to\infty} \hat W_\zeta^{i}=\hat W_\zeta$.
It turns out that $\big\langle\hat W_\zeta^{i\:2}\big\rangle^i$ shows a quasi-linear dependence in $\beta$ at low temperature. Thus, when injected into
(\ref{FiniteTemperatureSuperfluid}), the dependence in $\beta$ is canceled and the measured superfluid density becomes independent of the temperature.

Figure~\ref{ImprovedRhoS} shows the finite-temperature superfluid density as a function of $\beta$ and the improved estimator
for the zero-temperature superfluid density for the extended Bose-Hubbard model~(\ref{BoseHubbard}). As $\beta$ increases, the value given
by the improved estimator converges to the zero-temperature limit much faster than the finite-temperature superfluid density.
\begin{figure}[h]
  \centerline{\includegraphics[width=0.45\textwidth]{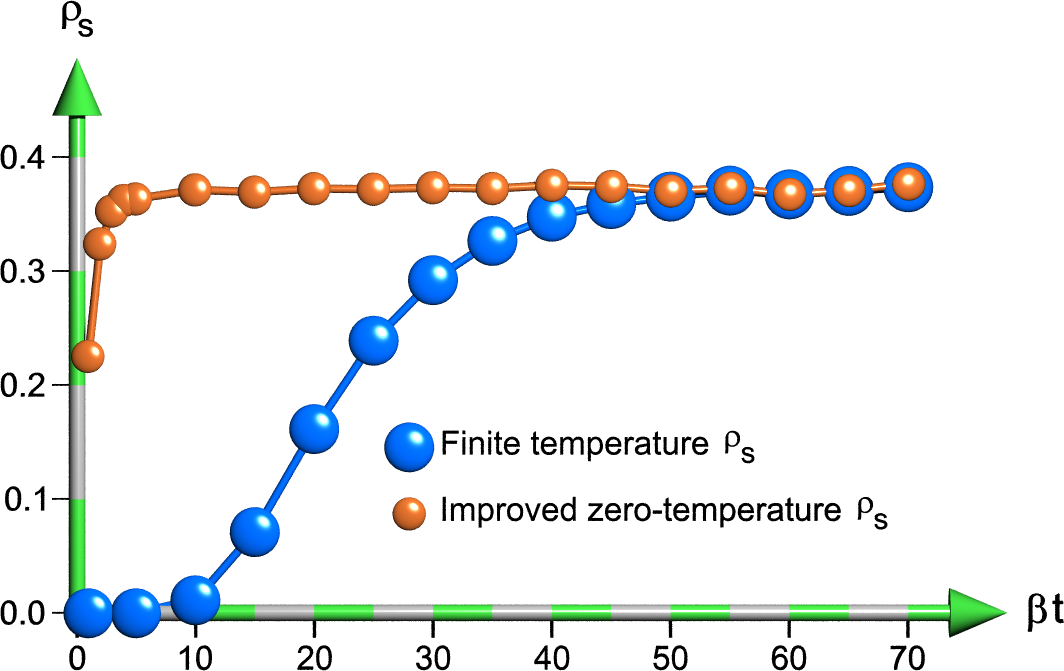}}
  \caption{(Color online) The finite-temperature superfluid density and the improved estimator of the zero-temperature superfluid
  density as functions of the inverse temperature $\beta$, for the extended Bose-Hubbard model~(\ref{BoseHubbard}). The lattice size is $L=100$,
  the onsite repulsion $U=14t$, the nearest neighbor interaction $V=0$, and the number of particles $N=50$. As $\beta$ increases, the value given
by the improved estimator converges to the zero-temperature limit much faster than the finite-temperature superfluid density.}
  \label{ImprovedRhoS}
\end{figure}

\subsubsection{\label{MeasDenDenCor}Imaginary time-dependent density-density correlation function}
Quantities like $\big\langle\stackrel{\leftarrow}{\textrm{\bf T}_\tau}\tilde n_{\vec r^{\:\prime}}(\tau^\prime)\tilde n_{\vec r}(\tau)\big\rangle^h=\big\langle\hat n_{\vec r^{\:\prime}}(\tau^\prime)\hat n_{\vec r}(\tau)\big\rangle^i$ can be measured directly in diagonal configurations,
since a given configuration $\mathcal C_n$ fully determines the values of $\hat n_{\vec r^{\:\prime}}(\tau^\prime)$ and $\hat n_{\vec r}(\tau)$. However,
in order to reduce the statistical fluctuations, we might want to take advantage of the invariance by imaginary time translation and
perform an average over the imaginary time axis. Also, if the Hamiltonian is invariant by space translation, the correlation function will
depend only on $\vec r^{\:\prime}-\vec r$, and fluctuations can be reduced further by summing over the space.

As a result we consider the correlation function
\begin{equation}
  \label{TimeCorrelation1} C_{\vec r}(\tau)=\frac{1}{L\beta}\sum_{\vec r^{\:\prime}}\int_0^\beta\big\langle\hat n_{\vec r+\vec r^{\:\prime}}(\tau+\tau^\prime)\hat n_{\vec r^{\:\prime}}(\tau^\prime)\big\rangle^i d\tau^\prime,
\end{equation}
which depends only on the translation $\vec r$ and time $\tau$ between the two correlated points. By introducing two identities~(\ref{Identity}),
we get
\begin{eqnarray}
  \label{TimeCorrelation2} C_{\vec r}(\tau) &=& \frac{1}{L\beta}\sum_{\vec r^{\:\prime}}\sum_{k=0}^n\sum_{q=0}^n\big\langle\hat n_{\vec r+\vec r^{\:\prime}}(\tau_q)\hat n_{\vec r^{\:\prime}}(\tau_k)\big\rangle_{\mathcal C_n}^i\\
  \nonumber                  &\times& \big[(M_1-m_1)\Theta(m_1<M_1)+(M_2-m_2)\Theta(m_2<M_2)\big],
\end{eqnarray}
with
\begin{equation}
  \left\lbrace
  \begin{array}{l}
    m_1=\max(\tau_k,\tau_q-\tau)\\
    m_2=\max(\tau_k,\tau_q+\beta-\tau)\\
    M_1=\min(\tau_{k+1},\tau_{q+1}-\tau)\\
    M_2=\min(\tau_{k+1},\tau_{q+1}+\beta-\tau)
  \end{array}
  \right..
\end{equation}
Again, Eq.~(\ref{TimeCorrelation2}) can be directly evaluated for any given configuration $\mathcal C_n$. However it should be
pointed that evaluating Eq.~(\ref{TimeCorrelation2}) for a single vector $\vec r$ and time $\tau$ is a process that has a
$L\beta^2$ complexity. If one is interested in getting full information on density-density correlations, it must be
evaluated for all possible vectors $\vec r$ and a number of time samples that is proportional to $\beta$. Thus the total complexity
of the measurement process becomes $L^2\beta^3$. We show in subsection~\ref{MeasDynamical} that it is actually possible to obtain full information
on density-density correlations with a~$L^2\beta+L\beta^2$ complexity by first evaluating correlations in the Fourier space, then performing an
invert Fourier transform.

\subsubsection{\label{MeasDynamical}Imaginary dynamical structure factor}
We define the imaginary dynamical structure factor as the space-time Fourier transform of the imaginary time-dependent density-density correlation function~(\ref{TimeCorrelation1}),
\begin{equation}
  \label{DynamicalStructureFactor} S(\vec k,\omega_p)=\frac{1}{L\beta}\sum_{\vec r}\int_0^\beta C_{\vec r}(\tau)e^{i\big(\vec k\cdot\vec r-\omega_p\tau\big)}d\tau
\end{equation}
with $\omega_p=\frac{2\pi p}{\beta}$. As discussed in subsection~\ref{MeasDenDenCor}, evaluating directly~(\ref{TimeCorrelation1})
for all vectors $\vec r$ and time $\tau$ has a $L^2\beta^3$ complexity, thus evaluating~(\ref{DynamicalStructureFactor}) for all
vectors $\vec k$ and frequencies $\omega_p$ may appear to be very difficult. Actually, this can be done with a $L^2\beta+L\beta^2$ complexity
by working directly in the $\omega$-space. For this purpose, we consider the Fourier transform of the imaginary time-dependent local
density operator $\hat n_{\vec r}(\tau)$, which can be directly evaluated in a configuration $\mathcal C_n$ by using~(\ref{DiagonalPart})
with $\hat\mathcal O_d=\hat n_{\vec r}$ and $f(\tau)=e^{-i\omega_p\tau}$,
\begin{eqnarray}
  \nonumber \big\langle\mathring n_{\vec r}(\omega_p)\big\rangle_{\mathcal C_n}^i &=&      \frac{1}{\beta}\big\langle\int_0^\beta \hat n_{\vec r}(\tau)e^{-i\omega_p\tau}d\tau\big\rangle_{\mathcal C_n}^i\\
  \nonumber                                                                       &=&      \frac{1}{\beta}\sum_{k=0}^n \big\langle\psi_k^{k+1}\big|\hat n_{\vec r}\big|\psi_k^{k+1}\big\rangle\\
  \label{FourierDensity}                                                          &\times& \frac{e^{-i\omega_p\tau_k}-e^{-i\omega_p\tau_{k+1}}}{i\omega_p}.
\end{eqnarray}
Evaluating (\ref{FourierDensity}) for all vectors $\vec r$ and frequencies $\omega_p$ is a process that
has a $L\beta^2$ complexity. Consider now the time Fourier transform of (\ref{TimeCorrelation1}):
\begin{eqnarray}
  \nonumber \mathring C_{\vec r}(\omega_p) &=& \frac{1}{\beta}\int_0^\beta C_{\vec r}(\tau)e^{-i\omega_p\tau}d\tau\\
  \label{FourierCorrelation}               &=& \frac{1}{L}\sum_{\vec r^{\:\prime}}\big\langle\mathring n_{\vec r^{\:\prime}}^\star(\omega_p)\mathring n_{\vec r+\vec r^{\:\prime}}(\omega_p)\big\rangle^i
\end{eqnarray}
Knowing $\mathring n_{\vec r}(\omega_p)$, we can obtain $\mathring C_{\vec r}(\omega_p)$ for all $\vec r$ and $\omega_p$ with an additional
$L^2\beta$ complexity. Finally, the imaginary dynamical structure factor can be obtained by performing a space Fourier transform,
\begin{equation}
  S(\vec k,\omega_p)=\frac{1}{L}\sum_{\vec r}\mathring C_{\vec r}(\omega_p)e^{i\vec k\cdot\vec r},
\end{equation}
which again has a $L^2\beta$ complexity. So the total complexity for measuring the dynamical structure factor is $L\beta^2+L^2\beta$.
In addition, the imaginary time-dependent density-density correlation function (\ref{TimeCorrelation1}) can be obtained from (\ref{FourierCorrelation}) for all vectors $\vec r$
and an arbitrary set of time indices by performing an invert time Fourier transform,
\begin{equation}
  C_{\vec r}(\tau)=\sum_p \mathring C_{\vec r}(\omega_p)e^{i\omega_p\tau},
\end{equation}
again with a total complexity $L\beta^2+L^2\beta$.

\subsubsection{\label{MeasGreen}Green functions and the momentum distribution function}
Green functions can be measured directly by using Eq.(\ref{ExpectationValue}). For example, the 4-point Green function
$G_{ijkl}=a_i^\dagger a_j^\dagger a_k^{\phantom\dagger} a_l^{\phantom\dagger}$ receives contributions from non-diagonal configurations
for which the matrix element $\big\langle L\big|G_{ijkl}/Q\big|R\big\rangle$ is non-zero. In the contributing configurations,
the denominator $\big\langle L\big|\hat\mathcal G\big|R\big\rangle$ is equal to $g(4)$, so the choice of the function $g(n)$
must be suitable for having a good frequency of measurement and a value of $g(4)$ that is not too small, in order to avoid strong
fluctuations. However $g(4)$ should not be too big, in order to have a good chance of generating diagonal configurations that
are needed for the normalization of the measurements.

The momentum distribution function, defined as $\mathring n(\vec k)=\big\langle\mathring a_{\vec k}^\dagger \mathring a_{\vec k}^{\phantom\dagger}\big\rangle$, measures the
average number of particles with momentum $\hbar\vec k$. For a $d$-dimensional lattice with $L$ sites, the creation and annihilation operators $\mathring a_{\vec k}^\dagger$ and $\mathring a_{\vec k}^{\phantom\dagger}$
of a particle with momentum $\vec k$ are defined as the Fourier transforms
\begin{eqnarray}
  \mathring a_{\vec k}^\dagger &=& \frac{1}{\sqrt{L}}\sum_p a_p^\dagger e^{i\vec k\cdot\vec r_p},\\
  \mathring a_{\vec k}^{\phantom\dagger} &=& \frac{1}{\sqrt{L}}\sum_q a_q^{\phantom\dagger} e^{-i\vec k\cdot\vec r_q},
\end{eqnarray}
where $a_p^\dagger$ and $a_q^{\phantom\dagger}$ are the creation and annihilation operators of
a particle on sites~$p$ and~$q$, respectively. The momentum distribution function takes the form
\begin{equation}
  \mathring n(\vec k)=\frac{1}{L}\sum_{p,q}\big\langle a_p^\dagger a_q^{\phantom\dagger}\big\rangle e^{i\vec k\cdot(\vec r_p-\vec r_q)},
\end{equation}
and can be obtained from 2-point Green functions, or the so-called \textit{one-particle density matrix}, $\rho_{pq}=\big\langle a_p^\dagger a_q^{\phantom\dagger}\big\rangle$.

\subsubsection{\label{MeasSpecific}The specific heat}
The specific heat $C_v$ is defined as the rate of change of the energy $E=\big\langle\hat\mathcal H\big\rangle$ with temperature $T$, $C_v=\frac{\partial E}{\partial T}\big|_V$,
keeping the volume (number of sites) constant.
A simple way to approximate $C_v$ consists in performing a numerical symmetric derivative of the energy, $C_v\approx\big(E(T+\delta T)-E(T-\delta T)\big)/2\delta T$.
This method works well, but requires to perform two simulations at temperatures $T-\delta T$ and $T+\delta T$ in addition to the simulation
at the temperature of interest $T$, and several attempts are usually needed in order to determine the ``best" value for $\delta T$.

It is actually possible to evaluate $C_v$ exactly with a single simulation, via the fluctuation-response theorem,
\begin{equation}
  \label{SpecificHeat} C_v=\frac{\partial\big\langle\hat\mathcal H\big\rangle}{\partial T}=\beta^2\big(\big\langle\hat\mathcal H^2\big\rangle-\big\langle\hat\mathcal H\big\rangle^2\big).
\end{equation}
In principle all terms in Eq.(\ref{SpecificHeat}) can be measured as described in this section. However this requires to build the list
of all terms present in $\hat\mathcal H^2$, whose number scales as $L^2$ for Hamiltonians with finite-range interactions ($L^4$ otherwise). In addition to being computationally
expensive, the sum of these terms may suffer from strong fluctuations. We show below that it is actually possible to evaluate
$\big\langle\hat\mathcal H^2\big\rangle$ globally by performing diagonal measurements only.

Since $\hat\mathcal H$ commutes with its exponential $e^{-\beta\hat\mathcal H}$, we can write
\begin{eqnarray}
  \nonumber \big\langle\hat\mathcal H^2\big\rangle &=& \big\langle\stackrel{\leftarrow}{\textrm{\bf T}_\tau}\big(\tilde\mathcal I(\hat\mathcal H)\big)^2\big\rangle^h\\
  \nonumber                                        &=& \big\langle\stackrel{\leftarrow}{\textrm{\bf T}_\tau}\big(\tilde\mathcal I(\hat\mathcal V)\big)^2\big\rangle^h+\sum_{k_1,k_2}\big\langle\stackrel{\leftarrow}{\textrm{\bf T}_\tau}\big[\tilde\mathcal I(\hat\mathcal T_{k_1})\tilde\mathcal I(\hat\mathcal T_{k_2})\big]\big\rangle^h\\
  \label{Hsquare}                                  &-& 2\sum_k\big\langle\stackrel{\leftarrow}{\textrm{\bf T}_\tau}\big[\tilde\mathcal I(\hat\mathcal V)\tilde\mathcal I(\hat\mathcal T_k)\big]\big\rangle^h
\end{eqnarray}
By applying the theorem (\ref{TheoremEquation}) for each term in~(\ref{Hsquare}), it follows that the specific heat can be measured as
\begin{eqnarray}
  \nonumber C_v &=& \big\langle n(n-1)\big\rangle^i+\big\langle\Big(\int_0^\beta\hat\mathcal V(\tau)d\tau\Big)^2\big\rangle^i\\
  \nonumber     &-& 2\big\langle n\int_0^\beta\hat\mathcal V(\tau)d\tau\big\rangle^i\\
                &-& \bigg(\big\langle\int_0^\beta\hat\mathcal V(\tau)d\tau-n\big\rangle^i\bigg)^2,
\end{eqnarray}
where $n$ is the total length of the operator string and the integrals are computed using~(\ref{DiagonalPart}) with $\hat\mathcal O_d=\hat\mathcal V$ and $f(\tau)=1$.

\subsubsection{\label{Entropy}The Entropy and the thermal susceptibility}
The statistical Von Neumann entropy is given by $\mathcal S=k\big(\ln\mathcal Z+\beta\big\langle\hat\mathcal H\big\rangle\big)$, where $k$ is
the Boltzmann constant. Because the actual value of the partition function $\mathcal Z$ is unknown in QMC simulations, a direct measurement of $\mathcal S$
is not possible. In order to overcome this problem, $\mathcal S$ is usually evaluated by performing a numerical integration of the specific heat $C_v$ over the temperature $T$,
\begin{equation}
  \mathcal S(T)=\int_0^T\frac{C_v}{T^\prime}dT^\prime,
\end{equation}
which requires a set of simulations at different temperatures.

We propose here an alternative which consists in performing a set of simulations in the grand-canonical ensemble (see section~\ref{GrandCanonical}) at different
values of the chemical potential. For simplicity, we consider here a Hamiltonian $\hat\mathcal H$ that describes identical particles,
for example Eq.(\ref{BoseHubbard}) or Eq.(\ref{HamiltonianExample}). The grand-canonical partition function takes the form
\begin{equation}
  \mathcal Z=\textrm{Tr }e^{-\beta\big(\hat\mathcal H-\mu\hat\mathcal N\big)},
\end{equation}
where $\hat\mathcal N$ is the operator that measures the total number of particles, and $\mu$ is the chemical potential.
Our thermodynamic control parameters are the temperature $T$ , the volume $V$ (number of sites $L$), and the chemical potential $\mu$.
We define the \textit{thermal susceptibility} $\chi_{th}$ as the rate of change of the average number of particles $N=\big\langle\hat\mathcal N\big\rangle$
with temperature $T$:
\begin{equation}
  \label{ThermalSusceptibility} \chi_{th}(T,V,\mu)=\frac{\partial N}{\partial T}\Big|_{V,\mu}
\end{equation}
By substituting $N=\frac{1}{\mathcal Z}\textrm{Tr }\hat\mathcal N e^{-\beta\big(\hat\mathcal H-\mu\hat\mathcal N\big)}$ in Eq.~(\ref{ThermalSusceptibility}),
and assuming that $\big[\hat\mathcal H,\hat\mathcal N\big]=0$, we get an expression for the thermal susceptibility that can be directly measured,
\begin{equation}
  \chi_{th}=\beta^2\Big[\big\langle\hat\mathcal N\big(\hat\mathcal H-\mu\hat\mathcal N\big)\big\rangle-\big\langle\hat\mathcal N\big\rangle\big\langle\hat\mathcal H-\mu\hat\mathcal N\big\rangle\Big],
\end{equation}
in a way similar to the energy as explained in~\ref{MeasEnergy}. Considering the energy $E=\big\langle\hat\mathcal H\big\rangle$ and the associated differential
$dE=Td\mathcal S-PdV+\mu dN$, where the pressure is defined as $P=-\frac{\partial E}{\partial V}\big|_{\mathcal S,N}$, and performing
a Legendre transformation over the variables $\mathcal S$ and $N$, we can define the grand-canonical potential $\Omega$ that depends
only on our natural variables, $\Omega(T,V,\mu)=E-T\mathcal S-\mu N=-PV$. Its differential takes the form
\begin{equation}
  d\Omega=-\mathcal SdT-PdV-Nd\mu.
\end{equation}
We can then extract a useful Maxwell relation,
\begin{equation}
  \frac{\partial\mathcal S}{\partial\mu}\Big|_{V,T}=\frac{\partial N}{\partial T}\Big|_{V,\mu},
\end{equation}
so the entropy can be easily obtained by integrating the
thermal susceptibility over the chemical potential and
keeping the temperature and the volume constant,
\begin{equation}
  \mathcal S(T,V,\mu)=\int_{\mu_0}^\mu\chi_{th}(T,V,\mu^\prime)d\mu^\prime,
\end{equation}
where $\mu_0$ is the critical value of the chemical potential
below which the average number of particles $N$ and the
thermal susceptibility $\chi_{th}$ are vanishing. Recently, this method has been applied successfully to Hamiltonians
describing diagonal and off-diagonal confinement~\cite{ODC}.

\section{\label{GrandCanonical}Simulation of the grand-canonical ensemble}
Consider a general Hamiltonian $\hat\mathcal H$ that describes $S$ species of particles, and let $\hat N_s$ be the number of particles of
a given species $s$ in an initial state. If it is possible to define $P$ charges $\hat\mathcal N_p$ that are conserved by the Hamiltonian,
\begin{equation}
  \left\lbrace
  \begin{array}{l}
    \hat\mathcal N_1=\gamma_{1,1}\hat N_1+\gamma_{1,2}\hat N_2+\cdots+\gamma_{1,S}\hat N_S\\
    \hat\mathcal N_2=\gamma_{2,1}\hat N_1+\gamma_{2,2}\hat N_2+\cdots+\gamma_{2,S}\hat N_S\\
    \cdots\\
    \hat\mathcal N_P=\gamma_{P,1}\hat N_1+\gamma_{P,2}\hat N_2+\cdots+\gamma_{P,S}\hat N_S,
  \end{array}
  \right.
\end{equation}
where $\gamma_{i,j}$ are integers, then the Hamiltonian commutes with $\hat\mathcal N_1,\hat\mathcal N_2,\cdots,\hat\mathcal N_P$, and the
\textit{canonical ensemble} is defined as the set of all states that contain exactly the same number of charges as the initial state.

By nature the SGF algorithm samples the canonical ensemble,
since all states are generated from an initial state by successive applications of $\hat\mathcal T$ operators. Simulating the canonical
ensemble can be convenient, especially for systems with several species~\cite{Feshbach,Spin1Bosons}. However it is sometimes useful to work in the
\textit{grand-canonical ensemble}~\cite{RousseauArXiv}, that is to say in the ensemble of states that contain any number of particles, especially for
magnetic systems. Thus, the best solution is to have an algorithm that can simulate both ensembles. We describe below a simple
extension that allows the SGF algorithm to simulate exactly the grand-canonical ensemble.

The idea is to add a non-conservative part $\hat\mathcal H_{nc}$ to the Hamiltonian,
\begin{equation}
  \hat\mathcal H_{nc}=\lambda\sum_{j,s}\big(a_{js}^\dagger+a_{js}^{\phantom\dagger}\big),
\end{equation}
where $j$ runs over all lattice sites and $s$ runs over all species, and $\lambda$ is an optimization parameter. This non-conservative
part allows the number of particles to fluctuate, so simulating the Hamiltonian $\hat\mathcal H^\prime=\hat\mathcal H+\hat\mathcal H_{nc}$ will sample
the grand-canonical ensemble. In order to make measurements that correspond to the actual Hamiltonian $\hat\mathcal H$, we can
perform a \textit{restricted} simulation of $\hat\mathcal H^\prime$ by applying the following conditions:
\begin{enumerate}
\item We allow at most one term of $\hat\mathcal H_{nc}$ at a time in the operator string.
\item We make measurements only if the operator string contains no $\hat\mathcal H_{nc}$ terms.
\end{enumerate}
The first condition is not required, but it allows to increase the probability to make a measurement. The second
condition ensures that the actual Hamiltonian $\hat\mathcal H$ is exactly simulated in the grand-canonical ensemble. Indeed, the extended
partition function $\mathcal Z^\prime(\beta,\tau)$ associated  with $\hat\mathcal H^\prime$ is a sum over configurations that
contain any number of $\hat\mathcal H_{nc}$ terms. Ignoring configurations
with $\hat\mathcal H_{nc}$ terms is equivalent to performing a renormalization of $\mathcal Z^\prime(\beta,\tau)$, such that the
resulting extended partition function $\mathcal Z(\beta,\tau)$ is only the sum of configurations that do not contain $\hat\mathcal H_{nc}$ terms.
As a result, configurations with no $\hat\mathcal H_{nc}$ terms are generated with the correct Boltzmann weight, and the grand-canonical
ensemble associated with the Hamiltonian $\hat\mathcal H$ is simulated exactly.

The value of $\lambda$ can be adjusted in order to tune the proportion of unphysical configurations with $\hat\mathcal H_{nc}$ terms.
$\lambda$ should be large enough to allow a fast decorrelation of the number of particles between different configurations. But
using a value too big reduces the probability of having a physical configuration and making a measurement. In practice we find that a good choice is $\lambda=1/L$, where
$L$ is the number of lattice sites.

Finally, the average number of particles $\big\langle \hat N_s\big\rangle$ of species $s$ can be adjusted via a chemical potential
$\mu_s$ by adding the term $-\mu_s\hat N_s$ to the Hamiltonian. An example of successful use of the SGF algorithm in both
canonical and grand-canonical ensembles is given in~\cite{RousseauArXiv}.

\section{Implementation}
\newcommand\psiTR{\big\langle\psi\big|\Kinetic\big|R\big\rangle}
\newcommand\LTpsi{\big\langle L\big|\Kinetic\big|\psi\big\rangle}
\newcommand\brapsi[1]{\left\langle #1\right|}
\newcommand\ketpsi[1]{\left| #1\right\rangle}
\newcommand\Green{\hat\mathcal G}
\newcommand\Kinetic{\hat\mathcal T}
\newcommand\GT{\brapsi{L} \Green \Kinetic \ketpsi{R}}
\newcommand\TG{\brapsi{L}\Kinetic \Green \ketpsi{R}}
Any implementation of the SGF algorithm will have to efficiently
evaluate all the update probabilities that appear in table~\ref{tab:summary_probabilities}.
In particular it will have to efficiently trace the values of the matrix elements $\GT$, $\TG$,
the left and right matrix elements $\psiTR$ and $\LTpsi$ and
the potential energy difference $\Delta V$. For a Hamiltonian with
arbitrary non-diagonal terms, for example with long range hopping, the only
general way of keeping track of these quantities is to calculate on the fly
those matrix elements. Therefore the cpu time of each update will scale
linearly with the number of non-diagonal terms in the Hamiltonian. This scaling
severely limits the system sizes that
are accessible with implementations that rely on recalculating all the probabilities at each update.

Fortunately, a major simplification is possible for the
most physically relevant Hamiltonians, for which the non-diagonal terms couple a constant number of sites,
as in the models of Eq.~(\ref{BoseHubbard}) and Eq.~(\ref{HamiltonianExample}).
In both these Hamiltonians the number of non-diagonal terms increases linearly 
in size and each non-diagonal term involves only a few neighboring site indices. 
In this section we will discuss an update algorithm where the cpu time per 
update scales logarithmically with the number of non-diagonal terms.

Without loss of generality we will consider the case with an insertion of a non-diagonal term to the right of $\Green$,
while the latter is moving to the left. 
The relative probability of the new state $\big|R^\prime\big\rangle$ after the insertion is
\begin{equation}
W^{LR}(R^\prime) = \brapsi{L} \Green \ketpsi{R^{\prime}} \brapsi{R^{\prime}} \Kinetic \ketpsi{R},
\label{eq:insertionweight}
\end{equation}
where the new matrix element of the green operator, $\brapsi{L} \Green \ketpsi{R^{\prime}}$,
is evaluated using Eq.~(\ref{GreenOperator}).
This expression depends only on the updated offset of the Green operator, that is to say the updated number of broken lines,
\begin{equation}
\label{eq:numbrokenlines} n_{L R^{\prime}}=\left|{\bf n}^{L}-{\bf n}^{R^{\prime}}\right|=\sum_{i}\left|n_{i}^{L}-n_{i}^{R^{\prime}}\right|,
\end{equation}
where the sum is over all sites $i$, and $n_{i}^{L}$ and $n_{i}^{R^{\prime}}$ are the corresponding 
occupancies of states $\ketpsi{L}$ and $\ketpsi{R^\prime}$ respectively. In most physical
models each term, $\Kinetic_k$, in $\Kinetic$ is a product of just a few creation/annihilation
operators, such as $a_{i}^{\dagger}a_{j}^{\phantom\dagger}$. Therefore the occupancies
of $\ketpsi{R^\prime}$ and $\ketpsi{R}$ will differ only over few
indices. We can separate out those indices in Eq.~(\ref{eq:numbrokenlines}) and
write the updated number of broken lines as
\begin{equation}
\label{eq:numbrokenlinesupdate} n_{L R^{\prime}}  =  n_{LR}+\delta n_{LR}\left[\Kinetic_k\right],
\end{equation}
where $\delta n_{L R}\left[\Kinetic_k\right]$ is the additional number 
of lines broken by $\Kinetic_k$ and is given by
\begin{equation}
\label{eq:offsetupdate} \delta n_{LR}\left[\Kinetic_k\right]=\sum_{i\text{ in }\Kinetic_k}\left(\left|n_{i}^{L}-n_{i}^{R}+\delta n_{i}(\Kinetic_k)\right|-\left|n_{i}^{L}-n_{i}^{R}\right|\right),
\end{equation}
where $n_{i}^{L}$ and $n_{i}^{R}$ are the occupancies of the $i^{th}$ index of $\ketpsi{L}$ and $\ketpsi{R}$ respectively,
$\delta n_{i}(\Kinetic_k)$ is the change in occupancy caused by $\Kinetic_k$ on index $i$,
and the summation extends only over the indices of the operator $\Kinetic_k$.
We will refer to $\delta n_{LR}\left[\Kinetic_k\right]$ as the 
``offset" of $\Kinetic_k$. Eq.~(\ref{eq:numbrokenlinesupdate}) and~(\ref{eq:offsetupdate})
suggest that the total number of broken lines can be updated by summing only over the 
usually few indices of $\Kinetic_k$ rather than all the indices of the configuration.

Furthermore we can rewrite Eq.~(\ref{eq:insertionweight}) as
\begin{equation}
W^{LR}(R^\prime)  =  g\left(n_{LR}+\delta n_{LR}\left[\Kinetic_k\right]\right)T_{k}^{R},
\end{equation}
where $T_{k}^{R}\equiv \brapsi{R^{\prime}} \Kinetic_k \ketpsi{R}$
is the matrix element of the term.
The offset is an integer that can take only few values, for example 
it can only take the values -2, 0, and +2 for the model~(\ref{BoseHubbard}) and the values -6, -4, -2, 0, +2, +4, and +6
for the model~(\ref{HamiltonianExample}).
It can therefore be used to categorize the non-diagonal terms.
Let $S\left(\delta n\right)$ be the group of non-diagonal terms
with the same offset, $\delta n$. For all those terms in $S\left(\delta n\right)$
the matrix element of the Green operator, $g\left(n_{LR}+\delta n\right)$,
is the same. 

The key idea is that the choice of an operator can be done in two steps, first
we choose an offset and then a term from the corresponding group.
The relative probability for choosing an offset is the sum of the relative
probabilities of all the terms with the same offset and can be written as
\begin{equation}
\label{eq:weightoffset} W^{LR}\left(\delta n\right)=g\left(n_{LR}+\delta n\right)\sum_{\Kinetic_k\text{ in }\Kinetic}T^{R}_k\delta_{\delta n,\delta n_{LR}\left(\Kinetic_k\right)},
\end{equation}
where the summation extends over all the non-diagonal terms and the Kr\"onecker delta
selects the terms with a particular offset $\delta n$. After the offset a non-diagonal term
is chosen from the group $S\left(\delta n\right)$ with relative probability $T^{R}_k$. 
The normalization of all the probabilities, $\GT$, reads
\begin{equation}
\label{eq:totalsum} \GT=\sum_{\delta n}W^{LR}\left(\delta n\right).
\end{equation}

So far we discussed about the insertion of a term $\Kinetic_k$ to the right. 
Removing a term from the right is equivalent to inserting $\Kinetic_k^{\dagger}$
as far as the configuration is concerned. Similarly an operator can
be added to the left which is the same as adding $\Kinetic_k^{\dagger}$
to the left or removed from the left which is equivalent to removing
$\Kinetic_k$.

When a term is added or removed, it will affect the offsets and the matrix
elements, $T^{R}_k$ of the terms that share an index with it.
An optimization occurs if instead of updating the matrix elements
and offsets of every term we update only the few affected terms.
In most physical models the number of 
non-diagonal terms that share any particular index does not change with 
the system size. For example if all non-diagonal terms are of the form 
$a_{i}^{\dagger}a_{j}^{\phantom\dagger}$, such as in the Bose-Hubbard model~(\ref{BoseHubbard}), then in $d$
dimensions, there are exactly $4d$ terms with any particular index. 
Therefore if such a term is inserted with indices $i$ and $j$ there 
are $8d-2$ terms that will be affected.

This observation leads to the development of a fast-update
algorithm for SGF. When a term is inserted or removed, a few non-diagonal terms
will need to be removed from the group of their original offset and
added to the group of their final offset. The pairs of $\left(\Kinetic_k,T^{R}_k\right)$
for each group can be stored in a binary search tree, one for each
offset $\delta n$ and direction of insertion relative to $\Green$. 
The binary search tree supports searching, inserting and removing of terms that scales logarithmically
in the number of terms. There are many potential implementations 
of such a binary search tree. One possibility is that the leaves correspond 
to the terms and their weight is their matrix element if they are members 
of the group or zero if they are not. The nodes have weights which are equal 
to the sum of the weights of their children. To choose a term one starts from
the root and selects one of the children randomly with their weight
as the relative probability and proceeds similarly until a leaf is
reached. When a term is inserted/removed, only the weights of the
nodes that are immediate ancestors need to be updated. Therefore
the logarithmic scaling is guaranteed. A simple picture of a binary
search tree with 4 terms is shown in Fig.~\ref{fig:A-binary-search-tree}.
\begin{figure}[h]
\begin{centering}
\centerline{\includegraphics[width=0.45\textwidth]{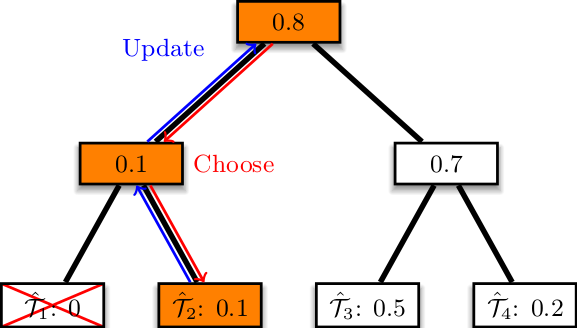}}
\end{centering}
\caption{A binary search tree for a model with 4 non-diagonal terms. 
There is one such tree for each group of terms with the same offset $\delta n$.
The leaves represent the terms and their weights, $T^{R}_k$. 
The nodes are the sum of the weights of all their children. 
In this example, the term $\Kinetic_{1}$ is not part of the group because it has zero weight. 
The colored boxes represent the nodes that will be accessed if term $\Kinetic_{2}$ is 
chosen, or when it is inserted, removed from the tree or its weight changes.
To choose an operator, one starts from the root and proceeds to the leaves,
by making a binary choice at every level with relative probabilities equal
to the stored weights of each node. When the weight of a term is updated
during an insertion, removal, or a change of occupancy, the change
is communicated to all its parents all the way up to the root. 
\label{fig:A-binary-search-tree}}
\end{figure}

\section{\label{Autocorrelation}Test of the algorithm}
In this section we illustrate the exactness of the SGF algorithm with global space-time update by making comparisons between QMC results and exact diagonalizations.
We also show that tuning the directionality of the new global space-time update allows to minimize the auto-correlation time.
Finally we show that the optimized auto-correlation time is smaller than the one obtained with the previous formulation of the
directed update~\cite{DirectedSGF}.

\subsection{Exactness of the SGF algorithm}
Figure \ref{ExactDiagonalization} shows a comparison between SGF results and an exact diagonalization for the non-trivial
model~(\ref{HamiltonianExample}) in the hard-core limit, for $2\times 2$ hexagons (12 sites), $\beta t=4$, at half-filling.
The total energy and the superfluid density are measured as explained in subsections~\ref{MeasEnergy} and~\ref{MeasSuperfluid}, respectively.
The agreement for the energy and the superfluid density illustrates the exactness of the SGF algorithm in both small and large $K/t$ limits.
\begin{figure}[h]
  \centerline{\includegraphics[width=0.45\textwidth]{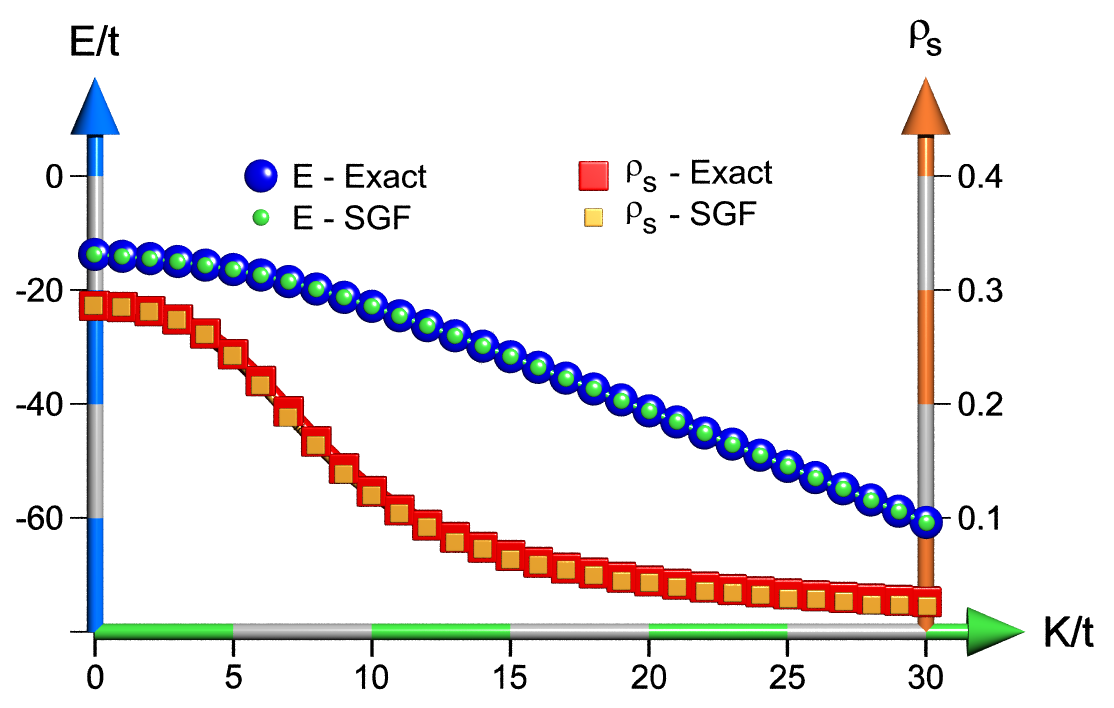}}
  \caption{(Color online) Comparison between the SGF algorithm with global space-time update and an exact diagonalization for the model with six-site ring-exchange
  terms in the hard-core limit. The agreement for the energy and the superfluid density is good in both small and large $K/t$ limits.}
  \label{ExactDiagonalization}
\end{figure}

The specific heat, measured as explained in subsection~\ref{MeasSpecific}, is shown on Fig.~\ref{CompSpecificHeat} as a function of
temperature $T$ for $K/t=5$ at half-filling. Again, the agreement is excellent both at low and at high temperature.
\begin{figure}[h]
  \centerline{\includegraphics[width=0.45\textwidth]{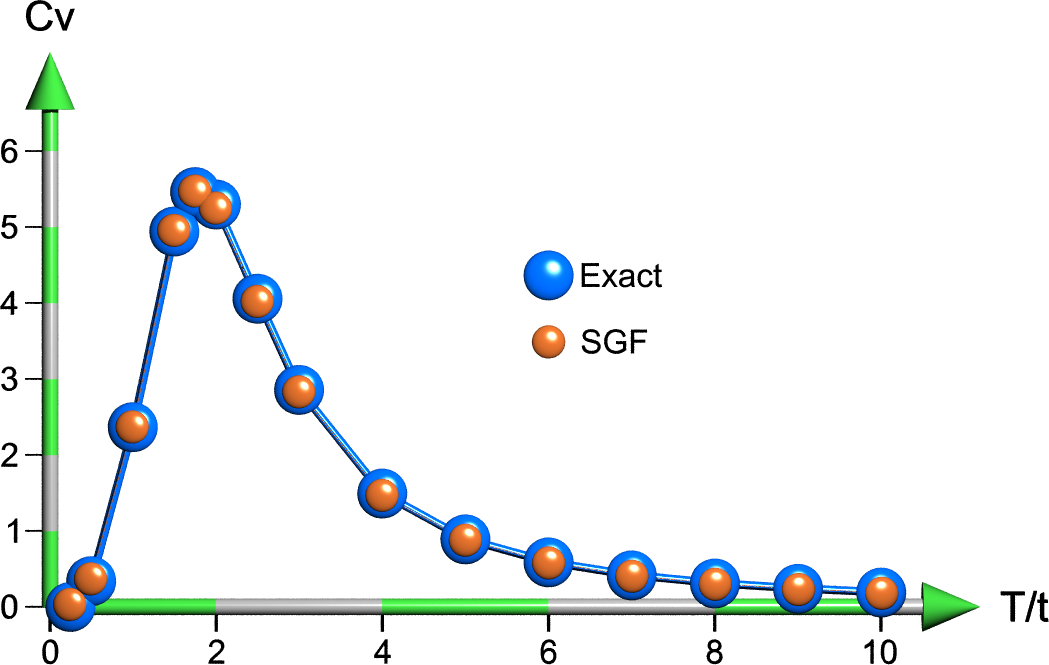}}
  \caption{(Color online) The specific heat $C_v$ as a function of temperature $T$. Comparison between the SGF algorithm with global space-time update
  and an exact diagonalization for the model with six-site ring-exchange
  terms in the hard-core limit. The two curves show a perfect agreement.}
  \label{CompSpecificHeat}
\end{figure}

\subsection{Auto-correlation time}
Let us consider a random variable $X$ and a set of $M$ successive samples $X(1),X(2),\cdots,X(M)$.
We define the auto-correlation function of $X$ at separation $\tau\in\mathbb{N}$ (shift in the sample indices) by
\begin{equation}
  C_X(\tau)=\frac{1}{M-\tau}\sum_{t=1}^{M-\tau}\big(X(t+\tau)-\big\langle X\big\rangle\big)\big(X(t)-\big\langle X\big\rangle\big),
\end{equation}
where $\big\langle X\big\rangle$ is the average value of $X$. If the samples $X(t)$ are independent from each other, then
$C_X(\tau)$ is vanishing for all $\tau>0$, namely $C_X(\tau)\propto\delta_{\tau,0}$. If the samples are correlated, then the auto-correlation
function is a decreasing function of $\tau$. The decay law (exponential, power, ...) depends on the system's Hamiltonian and
dimensionality, and whether one is close to a transition point or not. Thus we need a definition of the auto-correlation time
that is independent of these details. As a result, we define the auto-correlation time $\tau_c$ by the CPU time needed for the
auto-correlation function to drop by one order of magnitude with respect to its value at zero time separation:
\begin{equation}
  \label{AutocorrelationTime} C_X(\tau_c)=\frac{C_X(0)}{10}
\end{equation}

The auto-correlation time depends on the physical quantity $X$ that is measured. In the following we consider the model
(\ref{BoseHubbard}) with $L=100$ sites and the inverse temperature $\beta t=20$. We set the hopping parameter to $t=1$, and vary
the onsite repulsion $U$ and the first-nearest neighbor interaction $V$ in order to drive the system in a superfluid phase, a
Mott insulating phase, and a charge density wave phase. We calculate the auto-correlation
functions for the potential energy $E_p$, the kinetic energy $E_k$, the winding number $W$, and a non-trivial six-point correlation
function $C_{cor}^6$ defined by
\begin{eqnarray}
  \label{SixPoint} C_{cor}^6    &=& \Big\langle\stackrel{\leftarrow}{\textrm{T}_\tau}\big[\mathcal I_1\mathcal I_2\mathcal I_3\big]\Big\rangle^h\\
                   \mathcal I_1 &=& \frac{1}{\beta}\int\nolimits_0^\beta a_2^\dagger(\tau)a_3^{\phantom\dagger}(\tau)d\tau\\
                   \mathcal I_2 &=& \frac{1}{\beta}\int\nolimits_0^\beta a_3^\dagger(\tau)a_4^{\phantom\dagger}(\tau)d\tau\\
                   \mathcal I_3 &=& \frac{1}{\beta}\int\nolimits_0^\beta a_7^\dagger(\tau)a_6^{\phantom\dagger}(\tau)d\tau,
\end{eqnarray}
which is measured by making use of (\ref{TheoremEquation}).

As an illustration, Fig.~\ref{AutocorrelationMethod}
shows the normalized auto-correlation function of the six-point correlation function~(\ref{SixPoint})
as a function of CPU time for three different sets of parameters corresponding to a Mott insulator (top panel), a superfluid (middle panel),
and a charge density wave (bottom panel). The inset of the Mott insulator case with logarithmic-linear axes shows that the decay is nearly exponential,
while the inset of the charge density wave case with logarithmic-logarithmic axes shows a power law decay. This justifies our
definition of the auto-correlation time~(\ref{AutocorrelationTime}) which is independent of the decay law.
\begin{figure}[h]
  \centerline{\includegraphics[width=0.45\textwidth]{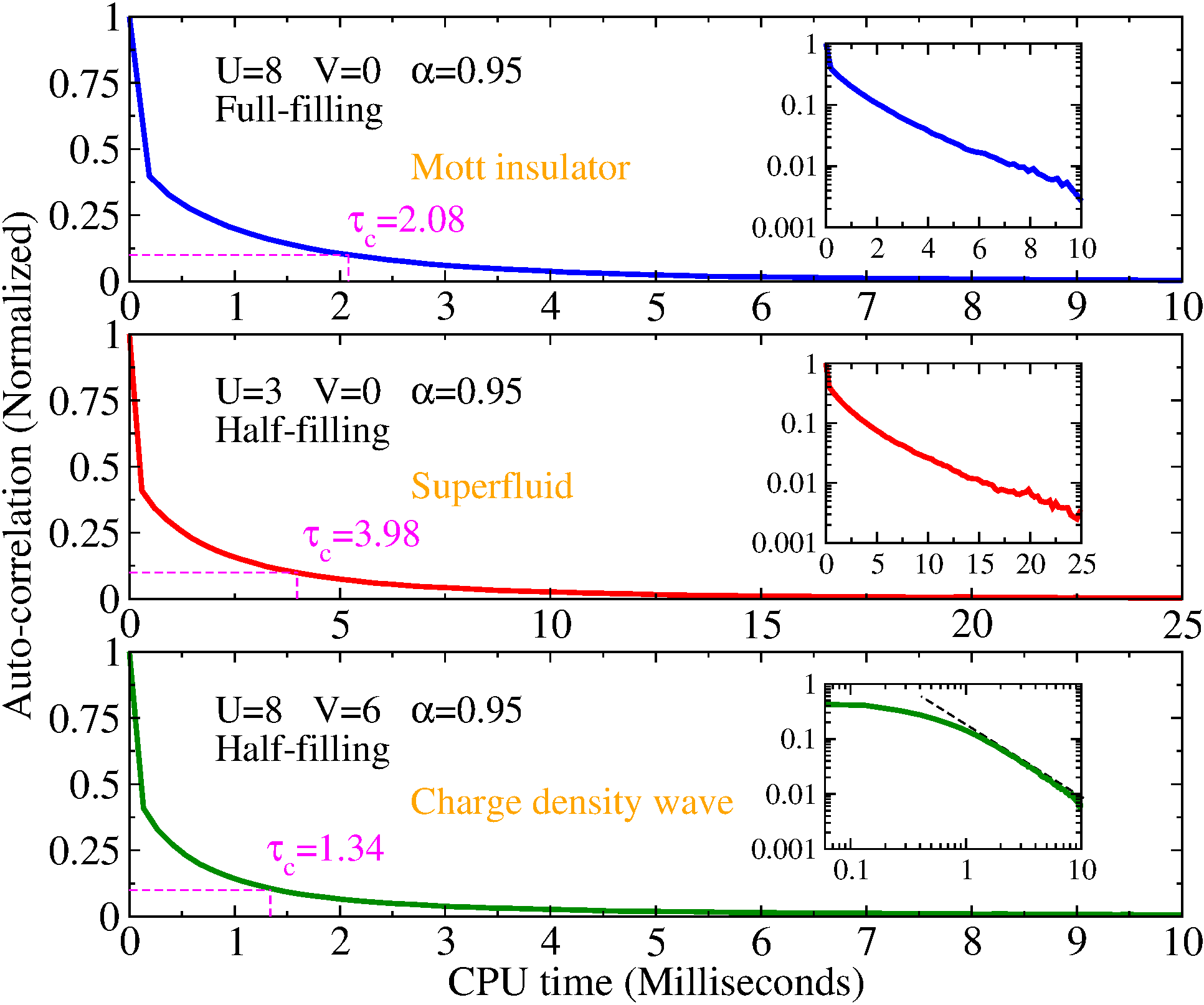}}
  \caption{(Color online) The normalized auto-correlation function of the six-point correlation function~(\ref{SixPoint}) as
  a function of CPU time for three different sets of parameters. For the Mott insulator case (top panel), the inset with
  logarithmic-linear axes reveals a nearly exponential decay. In the case of the charge density wave (bottom panel), the inset with logarithmic-logarithmic
  axes shows that the decay follows a power law.}
  \label{AutocorrelationMethod}
\end{figure}

\subsubsection{Effect of the directionality}
The parameter $\alpha$ allows us to control the directionality of the global space-time update, that it to say the average width
of the time window that is visited by the Green operator during a directed update. In other words, increasing $\alpha$ makes the
steps of the random walk of the Green operator in imaginary time larger, so the time needed to visit the full operator string
is smaller. But this also increases the CPU time needed to perform a directed update. The competition between these two effects
results in the existence of an optimal value of $\alpha$ that gives the smallest auto-correlation time for a given observable.

Figure~\ref{Directionality} shows the auto-correlation time of $E_p$, $E_k$, $W$ and $C_{cor}^6$ as functions of the directionality $\alpha$
for the superfluid phase (top panel), the Mott insulating phase (middle panel), and the charge density wave phase (bottom panel).
Note that for the Mott insulator and the charge density wave it is not possible to define an
auto-correlation time for the winding number $W$, because it is systematically vanishing for all samples. It is worth to
emphasize how the auto-correlation time is highly reduced for $C_{cor}^6$ from $\alpha=0$ to $\alpha=0.99$ by a factor of $\approx 20$
in the Mott insulating phase.
\begin{figure}[h]
  \centerline{\includegraphics[width=0.45\textwidth]{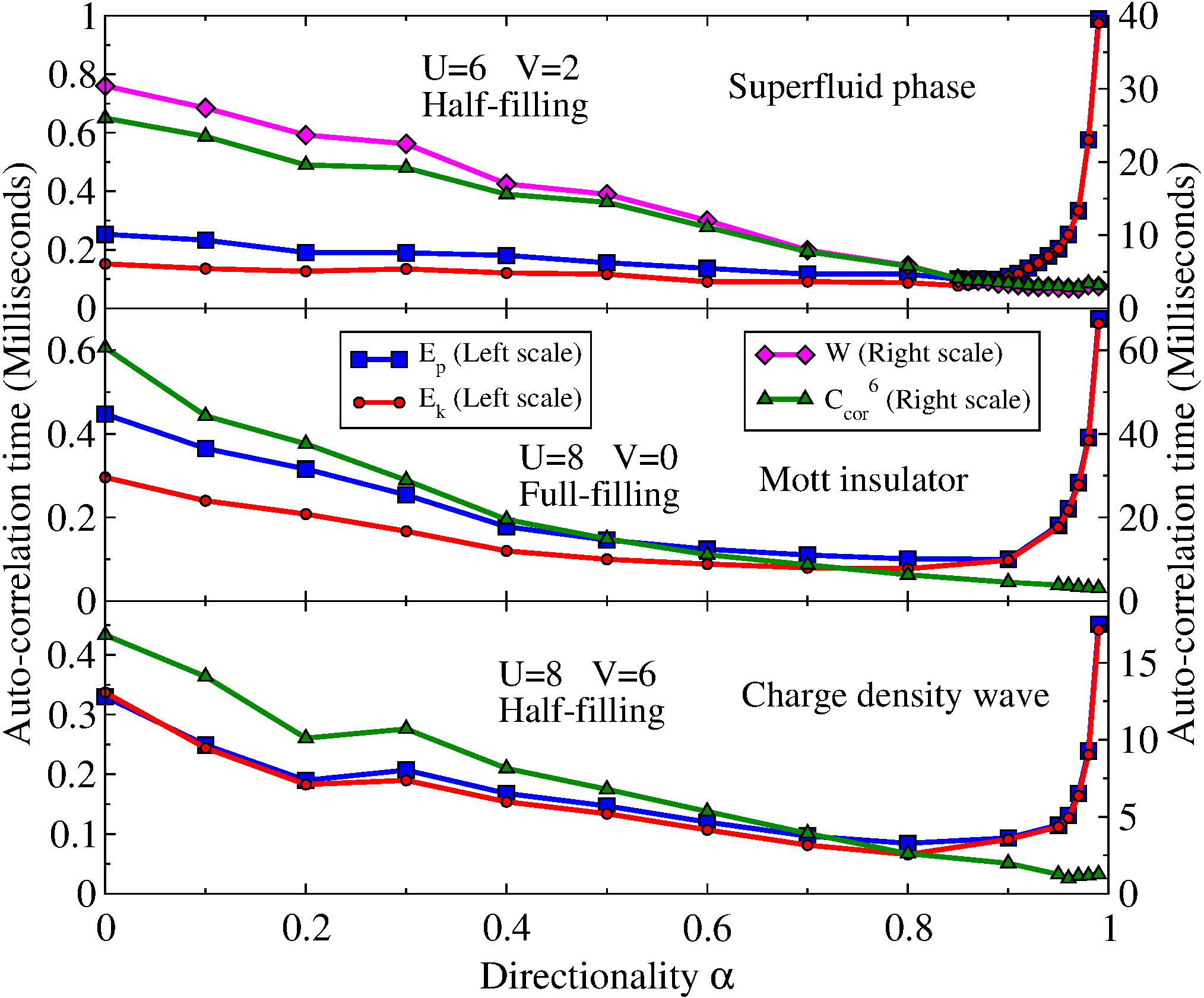}}
  \caption{(Color online) The auto-correlation time of $E_p$, $E_k$, $W$ and $C_{cor}^6$ as functions of the directionality $\alpha$
  in the superfluid phase (top panel), the Mott insulating phase (middle panel), and the charge density wave phase (bottom panel).}
  \label{Directionality}
\end{figure}

As a result, there exist an optimal value for $\alpha$ that gives the smallest auto-correlation time. But this optimal value
depends on the quantity that is measured and the phase in which the system is. In general we find that a good compromise is to
choose $\alpha\approx0.90$, which gives nearly optimized auto-correlation times for all observables in all phases of the system.

\subsubsection{Global space-time update versus previous formulation of directed update}
We compare here the auto-correlation times obtained with the global space-time update and those obtained with
the previous formulation of directed update~\cite{DirectedSGF}. The auto-correlation times given here are obtained by
using the optimal values of $\alpha$, which are determined graphically by using plots similar to Fig.~\ref{Directionality}
for each set of parameters.

We vary the values of the onsite potential $U$, the interaction between
nearest neighbors $V$, and the density of particles $\rho$. These parameters cover the superfluid phase, the Mott insulating
phase, and the charge density wave phase. Results for the potential energy $E_p$, the kinetic energy $E_k$,
the winding number $W$, and the correlation function $C_{cor}^6$ are shown on Fig.~\ref{Comparison} as scatter plots of the
auto-correlation times. One can notice that all points fall in the lower right part of the graphs, which indicates that the
auto-correlations times obtained with the global space-time update are smaller than those obtained with the previous directed
update in all cases.
\begin{figure}[h]
  \centerline{\includegraphics[width=0.45\textwidth]{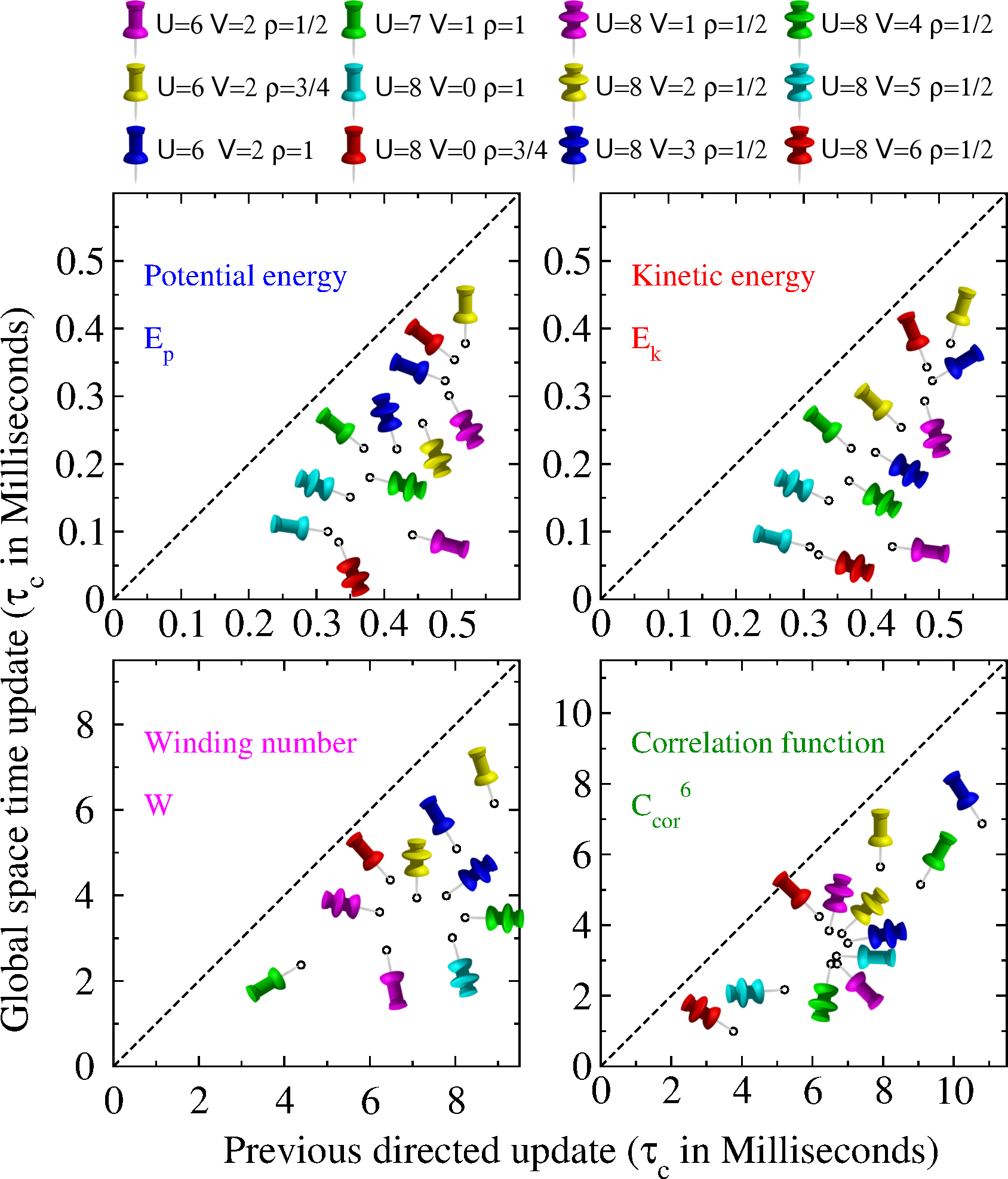}}
  \caption{(Color online) Comparison between the auto-correlation times obtained with the global space-time update and those
  obtained with the previous formulation of the directed update~\cite{DirectedSGF}. All points fall in the lower right part of the graphs, which indicates that the
  auto-correlations times obtained with the global space-time update are smaller than those obtained with the previous directed
  update in all cases.}
  \label{Comparison}
\end{figure}

\section{Conclusion}
We have presented the Stochastic Green Function (SGF) algorithm and showed that it is able to simulate any Hamiltonian that does not suffer
from the so-called ``sign problem".
We have proposed a new \textit{global space-time update} scheme which, in addition to being
\textit{directed}, has the advantage of reducing the auto-correlation time of the samples of measured quantities. The SGF algorithm
is the first quantum Monte Carlo (QMC) method that does not make any assumption on the form of the Hamiltonian. As a result, it can
be directly applied ``as is" to any Hamiltonian.
We have presented an optimized implementation where each update scales logarithmically with the system size, and which allows access to larger systems.
We have illustrated the capabilities of the SGF algorithm by applying it to a
Hamiltonian that includes six-site coupling terms, which is challenging for other QMC methods. In addition, we have shown that the SGF
algorithm can work in both canonical and grand-canonical ensembles. Finally, we have shown that various quantities of interest can
be measured by the algorithm, such as $n$-point Green functions, imaginary time-dependent correlation functions, the imaginary dynamical structure
factor, and the specific heat.

\appendix
\section{Proof of (\ref{HeisenbergInteraction}) and theorem (\ref{TheoremEquation})}
For convenience, we introduce the operator
\begin{equation}
 \label{OperatorU1}\hat\mathcal U(b,a)=\stackrel{\leftarrow}{\textrm{\bf T}_\tau}e^{\int_a^b\hat\mathcal T(\tau)d\tau},
\end{equation}
which satisfies the property
\begin{equation}
 \label{OperatorU2}\hat\mathcal U(c,b)\hat\mathcal U(b,a)=\hat\mathcal U(c,a)
\end{equation}
for $a\leq b\leq c$.
Using (\ref{InteractionPicture}) and (\ref{OperatorU1}), it follows that any set of arbitrary operators $\hat\mathcal A_1\cdots\hat\mathcal A_R$
satisfies
\begin{eqnarray}
  \nonumber e^{-\beta\hat\mathcal H}\tilde\mathcal A_R(\tau_R)\cdots\tilde\mathcal A_1(\tau_1) &=& e^{-\beta\hat\mathcal V}\hat\mathcal U(\beta,\tau_R)\hat\mathcal A_R(\tau_R)\\
  \nonumber                                                                                &\times& \hat\mathcal U(\tau_R,\tau_{R-1})\cdots\hat\mathcal U(\tau_2,\tau_1)\\
  \label{TheoremProof1}                                                                    &\times& \hat\mathcal A(\tau_1)\hat\mathcal U(\tau_1,0),
\end{eqnarray}
for any set of time-ordered indices $0<\tau_1<\cdots<\tau_R<\beta$.
By noticing that all $\tilde\mathcal A$, $\hat\mathcal A$, and $\tilde\mathcal U$ operators in~(\ref{TheoremProof1}) are written
in the chronological order, we can enclose them under a single time-ordering operator, which allows us to combine all $\tilde\mathcal U$ operators
into a single one. This leads to
\begin{eqnarray}
  \nonumber && e^{-\beta\hat\mathcal H}\stackrel{\leftarrow}{\textrm{\bf T}_\tau}\Big[\tilde\mathcal A_R(\tau_R)\cdots\tilde\mathcal A_1(\tau_1)\Big]\\
            && \quad\quad=e^{-\beta\hat\mathcal V}\stackrel{\leftarrow}{\textrm{\bf T}_\tau}\Big[\hat\mathcal A_R(\tau_R)\cdots\hat\mathcal A(\tau_1)e^{\int_0^\beta\hat\mathcal T(\tau)d\tau}\Big],
\end{eqnarray}
which is now valid for any set of unsorted time indices $\tau_1,\cdots,\tau_R\in[0;\beta[$. Taking the trace and normalizing
with $\mathcal Z$ leads to~(\ref{HeisenbergInteraction}). If in addition we integrate over all time indices, we get
\begin{eqnarray}
  \nonumber && \Big\langle\stackrel{\leftarrow}{\textrm{\bf T}_\tau}\Big[\tilde\mathcal I(\hat\mathcal A_R)\cdots\tilde\mathcal I(\hat\mathcal A_1)\Big]\Big\rangle^h\\
  \label{TheoremProof2} && \quad\quad=\frac{1}{\mathcal Z}\textrm{Tr }e^{-\beta\hat\mathcal V}\stackrel{\leftarrow}{\textrm{\bf T}_\tau}\Big[\hat\mathcal I(\hat\mathcal A_R)\cdots\hat\mathcal I(\hat\mathcal A_1)e^{\int_0^\beta\hat\mathcal T(\tau)d\tau}\Big].
\end{eqnarray}
Consider now the case where the operators $\hat\mathcal A_1\cdots\hat\mathcal A_R$ are all equal to the same non-diagonal Hamiltonian term $\hat\mathcal T_k$,
then~(\ref{TheoremProof2}) reduces to:
\begin{eqnarray}
  \nonumber && \Big\langle\stackrel{\leftarrow}{\textrm{\bf T}_\tau}\Big[\big(\tilde\mathcal I(\hat\mathcal T_k)\big)^R\Big]\Big\rangle^h\\
  \nonumber && \quad=\frac{1}{\mathcal Z}\textrm{Tr }e^{-\beta\hat\mathcal V}\stackrel{\leftarrow}{\textrm{\bf T}_\tau}\Big[\big(\hat\mathcal I(\hat\mathcal T_k)\big)^Re^{\int_0^\beta\hat\mathcal T(\tau)d\tau}\Big]\\
  \nonumber && \quad=\frac{1}{\mathcal Z}\textrm{Tr }e^{-\beta\hat\mathcal V}\stackrel{\leftarrow}{\textrm{\bf T}_\tau}\!\!\!\Bigg[e^{\int\limits_0^\beta\sum\limits_{q\neq k}\hat\mathcal T_q(\tau)d\tau}\sum_{n_k\geq 0}\frac{\bigg(\int\limits_0^\beta\hat\mathcal T_k(\tau)d\tau\bigg)^{n_k+R}}{n_k!\beta^R}\Bigg]\\
  \nonumber && \quad=\sum_{n_k\geq 0}\frac{n_k!}{\beta^R(n_k-R)!}\\
  \nonumber && \quad\times\underbrace{\frac{1}{\mathcal Z}\textrm{Tr }e^{-\beta\hat\mathcal V}\stackrel{\leftarrow}{\textrm{\bf T}_\tau}\!\!\!\Bigg[e^{\int\limits_0^\beta\sum\limits_{q\neq k}\hat\mathcal T_q(\tau)d\tau}\frac{\bigg(\int\limits_0^\beta\hat\mathcal T_k(\tau)d\tau\bigg)^n_k}{n_k!}\Bigg]}_{\begin{array}{c}\textrm{Probability of a string with}\\n_k\textrm{ operators of type }\hat\mathcal T_k\end{array}}\\
  \label{TheoremProof3} && \quad=\frac{1}{\beta^R}\Big\langle\frac{n_k!}{(n_k-R)!}\Big\rangle^i
\end{eqnarray}
It is easy to check that the above derivation is still valid when mixing up arbitrary operators $\hat\mathcal A_k$ with any
number of non-diagonal Hamiltonian terms $\hat\mathcal T_k$, hence~(\ref{TheoremEquation}) is proven.

\begin{acknowledgments}
VR is supported by 
NSF OISE-0952300 and DG by Lee Kuan Yew Fellowship of Singapore. This work used the Extreme Science and Engineering 
Discovery Environment (XSEDE), which is supported by National Science Foundation 
grant number DMR100007. We are grateful to Kalani Hettiarachchilage for useful discussions, and Mark Jarrell and Juana Moreno
for encouraging and supporting the development of the fast-update implementation.
\end{acknowledgments}

\end{document}